\documentclass[journal]{IEEEtran}
\newtheorem{definition}{Definition}
\newtheorem{lemma}{Lemma}
\newtheorem{theorem}{Theorem}
\newtheorem{corollary}{Corollary}
\newtheorem{proposition}{Proposition}

\usepackage{amsmath}
\usepackage{amssymb}
\usepackage{graphicx}
\usepackage{algorithm}
\usepackage{algpseudocode}
\algnewcommand{\IIf}[1]{\State\algorithmicif\ #1\ \algorithmicthen}
\algnewcommand{\EndIIf}{\unskip\ \algorithmicend\ \algorithmicif}
\ifCLASSINFOpdf
\else
\fi
\begin{document}
\title{Global Practical  Synchronization in Kuramoto Networks: A Submodular Optimization Framework}

\author{Andrew Clark,~\IEEEmembership{Member,~IEEE,}
        Basel Alomair,~\IEEEmembership{Member,~IEEE,}
        Linda Bushnell,~\IEEEmembership{Senior Member,~IEEE}
        and Radha Poovendran,~\IEEEmembership{Fellow,~IEEE}% <-this % stops a space
\thanks{A. Clark is with the Department of Electrical and Computer Engineering, Worcester Polytechnic Institute, Worcester, MA, 01609. {\tt aclark@wpi.edu}}%
\thanks{L. Bushnell and R. Poovendran are with the Department of Electrical Engineering, University of Washington, Seattle, WA, 98195-2500. {\tt\{lb2, rp3\}@uw.edu}}% <-this % stops a space
\thanks{B. Alomair is with the Center for Cybersecurity, King Abdulaziz City for Science and Technology, Riyadh, Saudi Arabia.   {\tt alomair@kacst.edu.sa}}}% <-this % stops a space

\maketitle

\begin{abstract}
Synchronization underlies  phenomena including memory and perception in the brain, coordinated motion of animal flocks, and stability of the power grid.  These synchronization phenomena are often modeled through networks of phase-coupled oscillating nodes.  Heterogeneity in the node dynamics, however, may prevent such networks from achieving the required level of synchronization.  In order to guarantee synchronization, external inputs can be used to pin a subset of nodes to a reference frequency, while the remaining nodes are steered toward synchronization via local coupling.  In this paper, we present a submodular optimization framework for selecting a set of nodes to act as external inputs in order to achieve  synchronization from almost any initial network state.  We derive  threshold-based sufficient conditions for synchronization, and then prove that these conditions are equivalent to connectivity of a class of augmented network graphs.  Based on this connection, we map the sufficient conditions for  synchronization to constraints on submodular functions, leading to efficient algorithms with provable optimality bounds for selecting input nodes.  We illustrate our approach via numerical studies of synchronization in networks from power systems, wireless networks, and neuronal networks.  %External inputs, including pacemakers in biological systems and anchor nodes in vehicle networks,   %Phase-coupled oscillator networks have been proposed to model synchronization phenomena.  In networks where a
\end{abstract}

\section{Introduction}
\label{sec:intro}
Synchronization plays a vital role in complex networks.  Stable operation of the power grid requires synchronization of buses and generators to a common frequency~\cite{dorfler2012synchronization}. Synchronized oscillations of neuronal firing provide a biological mechanism for aggregating information in perception \cite{eckhorn1988coherent} and memory \cite{klimesch1996memory}.  Coordinated motion of animals \cite{paley2007oscillator} occurs when a common heading is achieved.  The prevalence of synchronization across different application domains motivates the study of the basic principles underlying synchronization \cite{arenas2008synchronization}.

Phase-coupled oscillators have been proposed as a widely-applicable framework for studying synchronization \cite{rosenblum1996phase}.  In phase-coupled oscillator networks, the dynamics of each node's phase are determined by a diffusive coupling with its neighbors, together with an intrinsic frequency.  In particular, the Kuramoto model~\cite{acebron2005kuramoto}, which assumes sinusoidal coupling, has been extensively studied and is related to synchronization phenomena in power grids.  The sinusoidal coupling causes the node phases to approach synchronization, while the intrinsic frequency drives each node away from synchronization.

The existence and stability of synchronized states has been studied extensively in the  literature \cite{strogatz2000kuramoto,jadbabaie2004stability,chopra2009exponential}, including conditions based on the intrinsic frequencies, network topology, and degree of coupling between the nodes.    An important case is synchronization in the presence of external inputs \cite{childs2008stability}.  External inputs arise in applications including neuroscience, where they represent environmental stimuli \cite{eckhorn1988coherent} or deep brain stimulation \cite{mcintyre2010network}.  From an engineering standpoint, by introducing external inputs that pin a subset of nodes to a desired phase and frequency, a network that does not synchronize in the absence of inputs can be driven to a synchronized state, thus facilitating stability and performance of the network.

Existing analytical approaches to introducing external inputs  assume that the external input node is connected to all other nodes \cite{wang2013exponential}, or that the network has a specific topology such as a complete graph \cite{kori2004entrainment}.  Developing sufficient conditions for synchronization using external inputs in networks with arbitrary topology is an open problem.  Efficient algorithms for selecting a subset of input nodes in order to guarantee synchronization are also not available in the existing literature.

%While it has been shown that introducing external inputs facilitates synchronization, the effect of the choice of input nodes on the degree of synchronization has received little attention in the literature.  In particular, sufficient conditions for a set of input nodes to drive a network of phase-coupled oscillators to a desired level of synchronization from any initial state is an open problem.  Moreover, efficient algorithms for selecting input nodes that guarantee synchronization are currently lacking.

In this paper, we present an optimization framework for selecting a subset of phase-coupled oscillators to act as external inputs in order to guarantee  synchronization of a Kuramoto network.  We formulate  sufficient conditions for ensuring convergence of phase-coupled oscillators to a synchronized state from any given set of initial states.  We then develop a submodular optimization framework for selecting a minimum-size set of input nodes to achieve a desired level of synchronization, with provable bounds on the optimality of the chosen input nodes.  Our contributions are summarized as follows:
\begin{itemize}
\item We investigate two  synchronization problems.  The first problem is to ensure that  the phases of all of the oscillators converge to fixed points that are within a desired bound of a given reference phase (practical node synchronization).  The second problem is to guarantee that  the  phase differences of neighboring nodes converge to within a desired bound of each other (practical edge synchronization).  In both problems, the node frequencies must converge to the same value.
\item  We derive a set of sufficient conditions for a given set of input nodes to achieve practical node or edge synchronization from any given set of initial states.  We interpret our conditions as each node achieving a desired level of synchronization if a threshold number of neighbors reaches that level of synchronization.  %We present efficient algorithms for verifying these conditions, and demonstrate that we can trade off complexity of the algorithm and tightness of the synchronization criteria.
\item We develop a submodular optimization framework for selecting a minimum-size set of input nodes to achieve these synchronization conditions.  Our approach is to derive a connection between the threshold conditions and the connectivity of an augmented graph.  Based on this connection, we map our sufficient conditions to constraints on submodular functions.  We propose efficient algorithms for selecting input nodes to guarantee synchronization and analyze the optimality bounds of our algorithms.
\item We evaluate our approach through a numerical study of synchronization in  three classes of networks.  First, we consider synchronization of power grids using the IEEE 14 Bus test case \cite{14bus}.  Second, we study synchronization in geometric random graphs, motivated by vehicle coordination and wireless network synchronization problems. Finally, we investigate synchronization of neuronal networks based on the \emph{C. Elegans} dataset~\cite{varshney2011structural}.
\end{itemize} 
The rest of the paper is organized as follows.  In Section \ref{sec:related}, we give an overview of the related work.  Section \ref{sec:preliminaries} contains our system model and definitions of synchronization.  Sufficient conditions for synchronization are presented in Section \ref{sec:conditions}.  In Section \ref{sec:submodular}, we describe our submodular optimization approach to selecting input nodes that satisfy the sufficient conditions.  Section \ref{sec:simulation} contains our numerical study.  In Section \ref{sec:conclusion}, we conclude the paper and discuss directions for future work. 
\section{Related Work}
\label{sec:related}
The phase-coupled oscillator framework for modeling synchronization phenomena was introduced in the seminal work of Winfree \cite{winfree1967biological}.  Models of oscillation include Lorenz~\cite{jellison1996parameterization}, kick~\cite{mauroy2012kick}, Van der Pol~\cite{shinriki1981multimode}, and pulse-coupled~\cite{mirollo1990synchronization} oscillators.  The sinusoidally-coupled Kuramoto model was introduced in \cite{kuramoto1975self}.  Extensive studies have been performed on the mean-field behavior of the Kuramoto model with all-to-all coupling (i.e., each node is coupled to each other node) in the limit as the network size grows large \cite{acebron2005kuramoto}.  %Particular focuses of research have been the number of synchronized nodes, and conditions on the coupling coefficients between the nodes to ensure partial or full synchronization.

Synchronization of the Kuramoto model with a finite number of oscillators and arbitrary connection topology was studied in \cite{jadbabaie2004stability}.  The authors proved that convergence to the synchronized state is guaranteed when all nodes have identical intrinsic frequencies, and analyzed the feasibility and stability of synchronization under non-identical intrinsic frequencies.  Stable equilibria of the Kuramoto model in finite networks, including both synchronized and non-synchronized states, were analyzed in \cite{monzon2005global}.  More recently, the existence, uniqueness, and stability of partially synchronized states was studied in \cite{dorfler2013synchronization}, with application to power networks \cite{dorfler2012synchronization}.  These prior works considered synchronization in the absence of external inputs.

Synchronization in the presence of external inputs has achieved relatively less study. Numerical studies have estimated the region of attraction, defined as the set of initial states that converge to the desired state, of the Kuramoto model with inputs for the case of all-to-all coupling \cite{wiley2006size}.  Sufficient conditions for synchronization when there is a single input node that is connected to all other nodes were presented in \cite{wang2013exponential,popovych2011macroscopic,kori2004entrainment}.  These works do not, however, consider synchronization with external inputs in networks with arbitrary topology, and do not propose methods for selecting a subset of input nodes.

Steering a complex network to a desired state by pinning a set of nodes to a fixed value has been studied in the area of pinning control \cite{porfiri2008criteria,lu2009pinning,sorrentino2007controllability,yu2013synchronization}.  These existing works assume that each node's dynamics is a nonlinear function of the current state with linear coupling between the neighbors, and hence differs from our approach that considers nonlinear coupling between neighboring nodes.  To the best of our knowledge, conditions for pinning controllability of the Kuramoto model that we consider are not available in the existing literature.  %Furthermore, the approach taken in \cite{ employs the Master Stability Function formalism, which is fundamentally different from the threshold-based sufficient conditions that we derive.

%In the preliminary version of this paper~\cite{clark2015global}, a submodular optimization approach to selecting a minimum-size set of input nodes to guarantee that all node phases are within a desired bound of a reference phase was presented.  Selecting input nodes in order to ensure that the relative phase differences of neighboring nodes are within a desired bound, which is relevant to stability of power grids \cite{dorfler2012synchronization}, was not considered.

Selecting external input nodes to achieve synchronization can be viewed as part of the broader area of selecting input nodes to control complex networks.  Much recent work on selecting input nodes has focused on guaranteeing controllability of linear dynamics on the network \cite{liu2011controllability,ruths2014control}.  The assumption of linear dynamics, however, is not applicable to the nonlinear oscillators considered here.

Phase-coupled oscillators in general, and Kuramoto oscillators in particular, have found extensive applications.  Coupled oscillators were initially used to model natural phenomena such as bird flocking \cite{leonard2012decision} and fish schooling \cite{paley2007oscillator}.  At the level of individual cells, phase-coupled oscillators also provide a framework for heart pacemakers \cite{michaels1987mechanisms} and neuronal networks \cite{klimesch1996memory}. The prevalence of phase-coupled oscillators in nature has inspired engineering techniques for formation control \cite{paley2007oscillator} and time synchronization \cite{baldoni2010coupling}.  The phases of buses and generators in the power grid have also been modeled using phase-coupled oscillators, in order to understand whether the grid maintains a required level of synchronization for stability \cite{filatrella2008analysis}. 
\section{Model and Preliminaries}
\label{sec:preliminaries}
In this section, we describe the system model and oscillator dynamics.  We then define the notions of synchronization considered in this paper.  Finally, we give a preliminary result that will be needed for the proof of Theorem \ref{theorem:sufficient_practical_sync} in Section \ref{sec:conditions}.

\subsection{System Model}
\label{subsec:model}
A network of $n$ oscillators, indexed in the set $V = \{1,\ldots,n\}$ is considered (we use the term \emph{oscillator} and \emph{node} interchangeably, as the oscillators correspond to nodes in a graph).  Each oscillator $v \in V$ has a neighbor set $N(v) \subseteq V$, consisting of the set of oscillators that are coupled to $v$.  We assume that links are bidirectional, so that $u \in N(v)$ implies $v \in N(u)$.  An edge $(u,v)$ exists if $u \in N(v)$ and $v \in N(u)$.  We let $E$ denote the set of edges.  The graph $G=(V,E)$ is assumed to be connected; if not, our proposed input selection methods can be applied to each connected component of the graph. %, and let $p = |E|$.

Each oscillator $v$ has a time-varying phase $\theta_{v}(t)$.  The vector of oscillator phases at time $t$ is denoted $\boldsymbol{\theta}(t) \in \mathbb{R}^{n}$.  We assume that there are two types of oscillators, denoted \emph{input} and \emph{non-input} oscillators.  We let $A$ denote the set of input oscillators.  The phases of the non-input oscillators follow the Kuramoto dynamics~\cite{acebron2005kuramoto}
\begin{equation}
\label{eq:KM_follower}
\dot{\theta}_{v}(t) = -\sum_{u \in N(v)}{K_{uv}\sin{(\theta_{v}(t)-\theta_{u}(t))}} + \omega_{v}.
\end{equation}
In (\ref{eq:KM_follower}), the first term represents the coupling between the oscillators, while  $\omega_{v}$ is the intrinsic frequency of oscillator $v$ and describes the phase dynamics in the absence of coupling.  The coupling coefficient $K_{uv} > 0$ determines the relative strength of the two terms.  We assume throughout that the couplings between nodes are symmetric, so that $K_{uv} = K_{vu}$ for all $(u,v) \in E$.  %We define $\overline{K}_{v} \triangleq \sum_{u \in N(v)}{K_{uv}}$.

Each input oscillator $v \in A$ is assumed to be pinned to a desired frequency $\omega_{0}$ and phase offset $\theta_{0}$, so that $\dot{\theta}_{v}(t) = \omega_{0}$ and $\theta_{v}(t) = \omega_{0}t + \theta_{0}$ for all $v \in A$.  The overall oscillator dynamics are given by
\begin{equation}
\label{eq:KM}
\dot{\theta}_{v}(t) = \left\{
\begin{array}{ll}
-\sum_{u \in N(v)}{K_{uv}\sin{(\theta_{v}(t) - \theta_{u}(t))}} + \omega_{v}, & v \notin A \\
\omega_{0}, & v \in A
\end{array}
\right.
\end{equation}

We define a function $\sigma(x) : [-2\pi, 2\pi] \rightarrow [-\pi, \pi]$ as
\begin{displaymath}
\sigma(x) = \left\{
\begin{array}{ll}
x + 2\pi, & x \in [-2\pi, -\pi) \\
x, & x \in [-\pi, \pi) \\
x-2\pi, & x \in [\pi, 2\pi]
\end{array}
\right.
\end{displaymath}
The function $f(x)$ maps elements in $[-2\pi, 2\pi]$ into $[-\pi, \pi]$ while satisfying $\sin{(\sigma(x))} = \sin{(x)}$ and $\cos{(\sigma(x))} = \cos{(x)}$ for all $x$.  This function will be used to define the edge cohesiveness property in the following subsection.

\subsection{Definitions of Synchronization}
\label{subsec:sync}
We now define the notions of synchronization considered in this paper.  Analogous definitions for networks without external inputs are given in \cite{dorfler2013synchronization}.

The strongest form of synchronization is \emph{phase synchronization}, defined as follows.
\begin{definition}
\label{def:phase_sync}
The oscillators achieve \emph{phase synchronization} if there exists $\theta^{\ast}$ such that $\lim_{t \rightarrow \infty}{\theta_{v}(t)} = \theta^{\ast}$ for all $v \in V$.
\end{definition}

Phase synchronization is achieved if all oscillators converge to the same phase.  Since synchronization of all oscillators to the same phase is not possible in general \cite{jadbabaie2004stability}, weaker notions of synchronization have been proposed.  We first define \emph{node-} and \emph{edge-cohesiveness} as follows.

\begin{definition}[Node Cohesiveness]
A network of oscillators achieves $\gamma$-\emph{node cohesiveness} with parameter $\gamma \in [0, \frac{\pi}{4}]$ if there exists $T > 0$ such that, for all $t \geq T$, $|\theta_{v}(t) - (\omega_{0}t +\theta_{0})| < \gamma$.
\end{definition}

Node cohesiveness is achieved if all nodes converge to within a desired bound of the input node phase $(\omega_{0}t + \theta_{0})$.  The requirement that $\gamma \in [0,\frac{\pi}{4}]$ ensures that the relative phase differences between nodes are bounded by $\pi/2$, which is a requirement of synchronization in many real-world systems \cite{dorfler2012synchronization,dorfler2013synchronization}.

\begin{definition}[Edge Cohesiveness]
A network of oscillators achieves $\gamma$-\emph{edge cohesiveness} with parameter $\gamma \in [0, \frac{\pi}{2}]$ if there exists $T > 0$ such that, for all $t \geq T$ and all $(u,v) \in E$, $|\sigma(\theta_{v}-\theta_{u})| < \gamma$.
\end{definition}

  Edge cohesiveness implies that the relative differences between any pair of neighboring nodes is bounded above by $\gamma$.  In general, node cohesiveness is desirable in applications such as coordinated motion \cite{paley2007oscillator} and time synchronization \cite{baldoni2010coupling}, where all nodes must agree on a common phase.  Edge cohesiveness is desirable in applications including power systems \cite{dorfler2012synchronization}, where the relative differences between nodes must be within a certain range to ensure stability.

We observe that a network that is $\gamma$-node cohesive is $(2\gamma)$-edge cohesive.  Indeed, node cohesiveness implies that there exists $T > 0$ such that, for all $t \geq T$,
\begin{IEEEeqnarray*}{rCl}
|\theta_{v}(t) - \theta_{u}(t)| &=& |(\theta_{v}(t)-(\omega_{0}t + \theta_{0})) - (\theta_{u}(t)-(\omega_{0}t + \theta_{0}))|\\
 &\leq& |\theta_{v}(t) - (\omega_{0}t + \theta_{0})| + |\theta_{u}(t) - (\omega_{0}t + \theta_{0})| \\
  &<& \gamma + \gamma = 2\gamma.
\end{IEEEeqnarray*}

An additional synchronization notion is frequency synchronization, defined as follows.
\begin{definition}[Frequency Synchronization]
\label{def:freq_sync}
The oscillators achieve \emph{frequency synchronization} if there exists $\omega^{\ast}$ such that $\lim_{t \rightarrow \infty}{\dot{\theta}_{v}(t)} = \omega^{\ast}$ for all $v \in V$.
\end{definition}

Note that the only possible value for $\omega^{\ast}$ in Definition \ref{def:freq_sync} is the frequency of the input nodes, denoted $\omega_{0}$, since we have $\dot{\theta}_{v}(t) \equiv \omega_{0}$ for all $v \in A$ and $t > 0$.

We now define the main types of synchronization considered in this paper.
\begin{definition}[Practical Node and Edge Synchronization]
The oscillators achieve $\gamma$-\emph{practical node synchronization} if they achieve frequency synchronization and $\gamma$-node cohesiveness.  The oscillators achieve $\gamma$-\emph{practical edge synchronization} if they achieve frequency synchronization and $\gamma$-edge cohesiveness.
\end{definition}

The following lemma allows us to focus on the case where $\omega_{0} =\theta_{0} = 0$, so that all input oscillators have frequency and phase $0$.
\begin{lemma}
\label{lemma:frame_of_reference}
Define $\hat{\boldsymbol{\theta}}(t) = \boldsymbol{\theta}(t) - (\omega_{0}t + \theta_{0})\mathbf{1}$, where $\mathbf{1}$ denotes the vector of all $1$'s.  Then $\boldsymbol{\theta}(t)$ achieves $\gamma$-practical node (resp. edge) synchronization with frequency $\omega_{0}$ and input node phase $\theta_{0}$ if and only if $\hat{\boldsymbol{\theta}}(t)$ achieves $\gamma$-practical node (resp. edge) synchronization with frequency $0$ and input node phase $0$.
\end{lemma}

\begin{IEEEproof}
First, note that $\dot{\hat{\theta}}_{v}(t) = \dot{\theta}_{v}(t) - \omega_{0}$ for all $v \in V$.  Suppose that $\boldsymbol{\theta}(t)$ achieves $\gamma$-practical node synchronization with frequency $\omega_{0}$ and reference phase $\theta_{0}$.  Then for all $v \in V$,
\begin{displaymath}
\lim_{t \rightarrow \infty}{\dot{\hat{\theta}}_{v}(t)} = \lim_{t \rightarrow \infty}{\left(\dot{\theta}_{v}(t) - \omega_{0}\right)} = 0.
\end{displaymath}
Furthermore, there exists $T > 0$ such that, for all $t \geq T$ and $v \in V$, $$|\hat{\theta}_{v}(t)| = |\theta_{v}(t) - (\theta_{0} + \omega_{0}t)| \leq \gamma,$$ and hence $\gamma$-practical node synchronization is achieved.

Now, if $\boldsymbol{\theta}(t)$ achieves $\gamma$-practical edge synchronization with frequency $\omega_{0}$ and reference phase $\theta_{0}$, then there exists $T > 0$ such that, for all $t \geq T$ and $v \in V$, $$|\sigma(\hat{\theta}_{v} - \hat{\theta}_{u})| = |\sigma(\theta_{v}-\theta_{u})| \leq \gamma.$$
The proof of the converse is similar.
%The proof for $\gamma$-practical edge synchronization is similar, as is the proof of the converse.
\end{IEEEproof}

For $\gamma \in [0, \frac{\pi}{4}]$, define $\Lambda_{final} = \{||\boldsymbol{\theta}||_{\infty} \leq \gamma\}$.  Let $\Lambda_{init}$ be a set of feasible initial states, defined by $\Lambda_{init} = \{\boldsymbol{\theta} : |\theta_{i}| \leq \theta_{i}^{0}\}$ for some $\theta_{1}^{0},\ldots,\theta_{n}^{0}$.  Finally, let $\Lambda_{bound}$ be defined by $\Lambda_{bound} = \{\boldsymbol{\theta} : |\theta_{i}| \leq \theta_{i}^{max}\}$ for some $\theta_{1}^{max},\ldots,\theta_{n}^{max}$.
The desired condition that we consider for a set of input nodes to guarantee practical synchronization is defined as follows.
\begin{definition}
\label{def:input_node_goal}
Let $\Lambda_{final}$ and $\Lambda_{init}$ denote subsets of $[-\pi,\pi]^{n}$ with $\Lambda_{init} \subseteq \Lambda_{bound}$ and $\Lambda_{final} \cap \Lambda_{bound} \neq \emptyset$.  A set of input nodes is said to \emph{guarantee practical synchronization} if  for any $\boldsymbol{\theta}(0) \in \Lambda_{init}$, (i) $\lim_{t \rightarrow \infty}{\boldsymbol{\theta}(t)} = \boldsymbol{\theta}^{\ast}$ for some $\boldsymbol{\theta}^{\ast} \in \Lambda_{final}$, and (ii) $\boldsymbol{\theta}(t) \in \Lambda_{bound}$ for all $t \geq 0$.
\end{definition}

In Definition \ref{def:input_node_goal}, the condition (i) implies that the oscillators eventually achieve frequency synchronization at an equilibrium point that lies within a desired region $\Lambda_{final}$.  Condition (ii) implies that, prior to convergence, the oscillators remain within a desired ``safe'' region $\Lambda_{bound}$.  As an example, in a power system, the goal may be to ensure that a set of generators reach frequency synchronization with relative phase differences between the nodes within a desired region $\Lambda_{final}$, while ensuring that $$\boldsymbol{\theta}(t) \in \Lambda_{bound} \triangleq \left\{\boldsymbol{\theta} : |\theta_{v}-\theta_{u}| \leq \frac{\pi}{2} \ \forall (u,v) \in E\right\}$$ for all $t \geq 0$.  If this condition does not hold, generators will lose synchronism with respect to each other, and one or more generators may trip to avoid hardware damage, potentially leading to grid instability.

Finally, we define a metric that describes the worst-case distance of each oscillator from the input node phase at equilibrium. \begin{definition}
\label{def:input_cohesiveness}
Suppose that a set of oscillators achieves frequency synchronization for any $\boldsymbol{\theta}(0) \in \Lambda_{init}$.  The \emph{input cohesiveness} is defined by
\begin{equation}
\label{eq:input_coh}
g(A) = \max{\{||\boldsymbol{\theta}^{\ast}||_{\infty} : \boldsymbol{\theta}^{\ast} \in \Theta\}}
\end{equation}
where $$\Theta = \{\boldsymbol{\theta} \in \mathbb{R}^{n} : \lim_{t \rightarrow \infty}{\boldsymbol{\theta}(t)} = \boldsymbol{\theta}^{\ast} \ \mbox{for some $\boldsymbol{\theta}(0) \in \Lambda_{init}$}\}.$$
\end{definition}
The metric $g(A)$ corresponds to the smallest $\gamma$ such that the oscillators achieve $\gamma$-practical node synchronization from input set $A$.

\subsection{Preliminary Result}
\label{subsec:consensus_result}
The following preliminary result of~\cite{moreau2004stability} will be needed  in Section \ref{sec:conditions}.  First, for any $\delta > 0$ and any matrix $B$, the $\delta$-digraph is defined as the digraph where edge $(i,j)$ exists if $B_{ij} \geq \delta$.

\begin{theorem}[\cite{moreau2004stability}, Theorem 1]
\label{theorem:consensus}
Consider the linear system $\dot{x}(t) = F(t)x(t)$, where $F(t)$ is a time-varying system matrix.  Assume that the system matrix is a bounded piecewise continuous function of time, and that for every time $t$ the system matrix is Metzler (i.e., all off-diagonal elements are nonnegative) and has zero row sums. Suppose further that there is an index $k \in \{1,\ldots,n\}$, a threshold value $\delta > 0$, and an interval length $T > 0$ such that for all $t \in \mathbb{R}$ the $\delta$-digraph associated to $$\int_{t}^{t + T}{F(s) \ ds}$$ has the property that all nodes may be reached from the node $k$. Then the set of states $\{x^{\ast}\mathbf{1} : x^{\ast} \in \mathbb{R}\}$ is uniformly exponentially stable.  In particular, all components of any solution converge to a common value as $t \rightarrow \infty$.
\end{theorem}

The theorem gives a sufficient condition for a linear system with time-varying weights to converge to consensus.  Theorem \ref{theorem:consensus} will be used to prove that, under certain conditions on the oscillator phases, the frequencies $\{\dot{\theta}_{v}(t) : v \in V\}$ converge to the frequencies of the input nodes. 
\section{Sufficient Conditions for Practical Synchronization}
\label{sec:conditions}
In this section, we formulate the problem of selecting a set of inputs to guarantee practical synchronization and provide sufficient conditions for a set of inputs to achieve practical synchronization.  We first show that the oscillators achieve practical synchronization if there exists a positive invariant set containing the set of possible initial states of the oscillators, and if any initial state in the invariant set will eventually converge to the set of admissible final states.  We then derive sufficient conditions for existence of the positive invariant set, as well as convergence to practical synchronization.
%In this section, we provide sufficient conditions for a set of inputs to achieve practical synchronization.  We first show that the oscillators achieve practical node (resp. edge) synchronization if there exists at least one stable equilibrium point satisfying node (resp. edge) cohesiveness, and no stable equilibria that do not satisfy node (resp. edge) cohesiveness.  We then formulate sufficient conditions for existence of a stable and cohesive equilibrium point, as well sufficient conditions for non-existence of stable equilibria that do not satisfy cohesiveness.  We provide efficient algorithms for verifying the sufficient conditions for a given set of inputs.

\subsection{Statement of Sufficient Condition}
\label{subsec:equilibrium}

The following theorem gives a sufficient condition for a set of oscillators to guarantee practical synchronization.
\begin{theorem}
\label{theorem:sufficient_practical_sync}
Suppose that there exists a set $\Lambda_{PI}$ such that the following conditions hold: (a) $\Lambda_{init} \subseteq \Lambda_{PI} \subseteq \Lambda_{bound}$, (b) $\Lambda_{PI}$ is \emph{positive invariant}, i.e., if $\boldsymbol{\theta}(0) \in \Lambda_{PI}$, then $\boldsymbol{\theta}(t) \in \Lambda_{PI}$ for all $t\geq 0$, and (c) If $\boldsymbol{\theta}(0) \in \Lambda_{PI}$, then there exists $T$ such that $t \geq T$ implies $\boldsymbol{\theta}(t) \in \Lambda_{final}$.  Then the set of input nodes $A$ satisfies conditions (i) and (ii) of Definition \ref{def:input_node_goal} and hence guarantees practical synchronization.
\end{theorem}

\begin{IEEEproof}
In order to prove condition (i) of Definition \ref{def:input_node_goal}, consider $\dot{\boldsymbol{\theta}}(t)$, which has dynamics
 \begin{equation}
 \label{eq:theta_dot_dynamics}
 \ddot{\theta}_{v}(t) = -\sum_{u \in N(v)}{\left[K_{uv}\cos{(\theta_{v}(t) - \theta_{u}(t))}(\dot{\theta}_{v}(t) - \dot{\theta}_{u}(t))\right]}
 \end{equation}
 for $v \notin A$ and $\ddot{\theta}_{v}(t) \equiv 0$ for $v \in A$.  We now define dynamics of the form in Theorem \ref{theorem:consensus} in order to analyze the convergence of (\ref{eq:theta_dot_dynamics}).  Let $x(t) \in \mathbb{R}^{n+1}$ denote the state variable, where $x_{n+1}$ is the state of a ``super node'' with dynamics $\dot{x}_{n+1}(t) \equiv 0$.  Define the system matrix $F(t)$ by
 \begin{displaymath}
 F_{vu}(t) = \left\{
 \begin{array}{l}
 \cos{K_{uv}(\theta_{v}(t) - \theta_{u}(t))}, \ \mbox{for } (u,v) \in E, v \notin A\\
 %& i \notin A \\
 -\sum_{s \in N(v)}{K_{sv}\cos{(\theta_{v}(t)-\theta_{s}(t))}}, \  u=v,  u \notin A \\
 -1, \ \mbox{for } u=v, u \in A \\
 1, \ \mbox{for }  v \in A, u = (n+1) \\
  0, \ \mbox{for } v = n+1
  \end{array}
  \right.
  \end{displaymath}
  By  condition (c), there exists $T > 0$ such that $|\theta_{v}(t) - \theta_{u}(t)| < \pi/2$ for all $t > T$.  Hence $F(t)$ is a bounded, piecewise continuous Metzler matrix with rows that sum to zero, and the connectivity of the graph $G$ implies that node $(n+1)$ is connected to all other nodes in the associated $\delta$-digraph.  By Theorem \ref{theorem:consensus}, $x(t)$ converges to a state $x^{\ast}\mathbf{1}$.

  Now, if we set $x_{v}(0) = x_{n+1}(0) = 0$ for all $v \in A$ and $x_{v}(0) = \dot{\theta}_{v}(0)$ for all $v \notin A$, then the trajectory of $[x_{1}(t) \cdots x_{n}(t)]^{T}$ will be identical to the trajectory of $\dot{\boldsymbol{\theta}}(t)$.  Hence, by Theorem \ref{theorem:consensus}, $\lim_{t \rightarrow \infty}{\dot{\boldsymbol{\theta}}(t)} = \omega^{\ast}\mathbf{1}$ for some $\omega^{\ast} \in \mathbb{R}$.  Moreover, since $\dot{\theta}_{v}(t) \equiv 0$ for all $v \in A$, we must have $\omega^{\ast} = 0$, implying frequency synchronization.

  Since $\dot{\boldsymbol{\theta}}$ converges to $0$, $\boldsymbol{\theta}(t)$ converges to a fixed point $\boldsymbol{\theta}^{\ast}$.  By condition (c), since $\boldsymbol{\theta}(t) \in \Lambda_{final}$ for $t$ sufficiently large, $\boldsymbol{\theta}^{\ast} \in \Lambda_{final}$.  This completes the proof of condition (i) of Definition \ref{def:input_node_goal}.

  The proof of condition (ii) in Definition \ref{def:input_node_goal} follows from (a) and (b), which imply that $\boldsymbol{\theta}(t) \in \Lambda_{PI} \subseteq \Lambda_{bound}$ for all $t \geq 0$ whenever $\boldsymbol{\theta}(0) \in \Lambda_{init}$.
\end{IEEEproof}

\begin{figure}[!ht]
\centering
\includegraphics[width=3in]{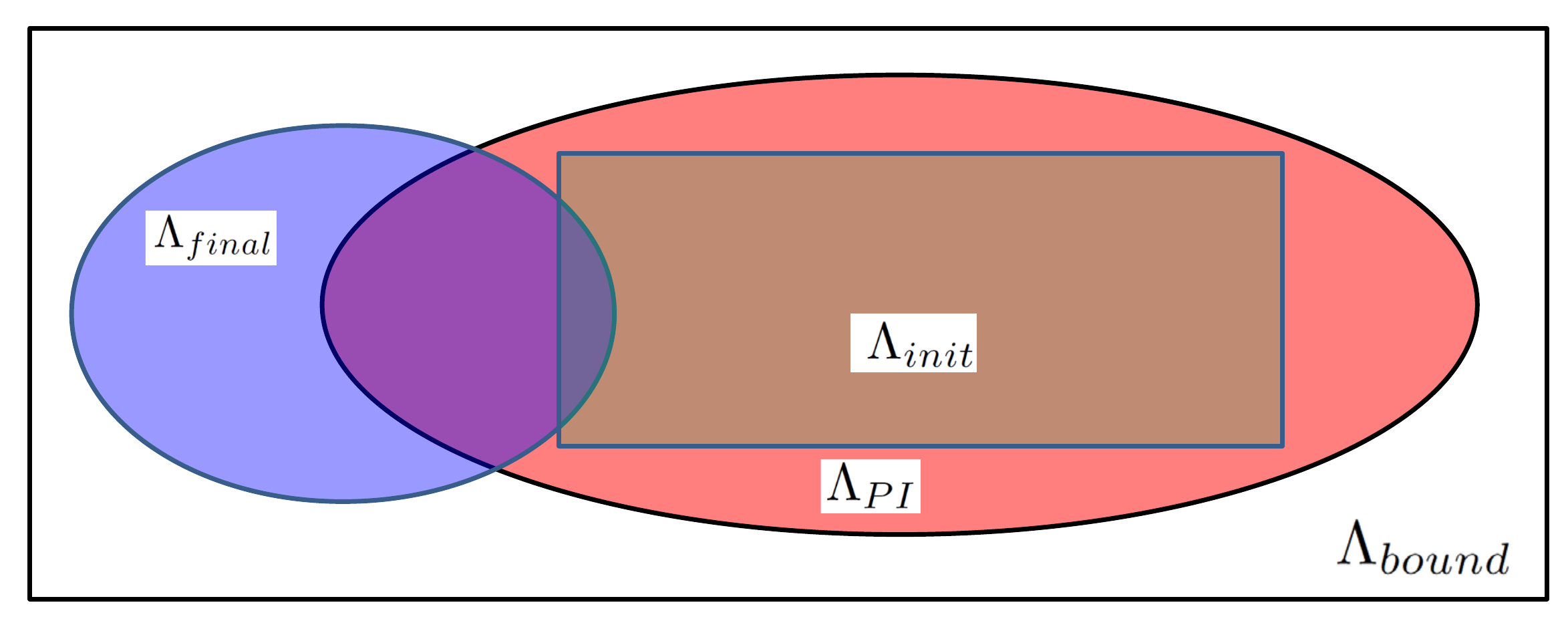}
\caption{Illustration of the sets $\Lambda_{bound}$, $\Lambda_{init}$, $\Lambda_{final}$, and $\Lambda_{PI}$.}
\label{fig:set_illustration}
\end{figure}

Figure \ref{fig:set_illustration} illustrates the relationship between the sets $\Lambda_{bound}$, $\Lambda_{init}$, $\Lambda_{final}$, and $\Lambda_{PI}$ of Definition \ref{def:input_node_goal} and Theorem \ref{theorem:sufficient_practical_sync}.

%\begin{proposition}
%\label{prop:cohesiveness_well_defined}
%If Conditions (a) and (b) of Theorem \ref{theorem:sufficient_practical_sync} hold, then

%Satisfying the conditions of Theorem \ref{theorem:sufficient_practical_sync} relies on the existence of  a set $\Lambda_{PI}$ that satisfies (a)--(c).
We observe that, if conditions (a) and (b) of Theorem \ref{theorem:sufficient_practical_sync} hold, then the oscillators achieve frequency synchronization and the input cohesiveness metric $g(A)$ is well-defined. The following section describes a procedure for testing whether a positive invariant set $\Lambda_{PI}$ satisfying the conditions (a)--(c) of Theorem \ref{theorem:sufficient_practical_sync} exists for a given set of input nodes $A$.

\subsection{Identifying a Positive Invariant Set}
\label{subsec:conditions_PI}

In this section, we develop a procedure for identifying a subset $\Lambda_{PI}$ that is positive invariant and satisfies $\Lambda_{init} \subseteq \Lambda_{PI} \subseteq \Lambda_{bound}$, for a given set of input nodes $A$, provided that such a positive invariant set exists.  This corresponds to conditions (a) and (b) of Theorem \ref{theorem:sufficient_practical_sync}.  We first state the following proposition that gives a sufficient condition for positive invariance.

\begin{proposition}
\label{prop:PI_sufficient_condition}
Suppose that a set $\Lambda_{PI}$ is defined by $\Lambda_{PI} = \{\boldsymbol{\theta} : |\theta_{v}| \leq \overline{\theta}_{v}\}$ for some $\{\overline{\theta}_{v} : v \in V\}$.  If
\begin{equation}
\label{eq:PI_condition}
\sum_{u \in N(v)}{\min{\left\{K_{uv}\sin{(\overline{\theta}_{v}-\theta_{u})} : |\theta_{u}| \leq \overline{\theta}_{u}\right\}}} > |\omega_{v}| + \epsilon
\end{equation}
 for all $v \in V$ and some $\epsilon > 0$, then the set $\Lambda_{PI}$ is positive invariant.
\end{proposition}

\begin{IEEEproof}
Suppose that (\ref{eq:PI_condition}) holds for all $v \in V$ and yet $\Lambda_{PI}$ is not positive invariant.  Let $\boldsymbol{\theta}(0) \in \Lambda_{PI}$ be such that $\boldsymbol{\theta}(t) \notin \Lambda_{PI}$ for some $t$, and define $$t^{\ast} \triangleq \inf{\{t: \theta_{v}(t) > \overline{\theta}_{v} \ \mbox{for some v}\}}.$$  Then $|\theta_{v}(t^{\ast})| = \overline{\theta}$ for some $v \in V$, and we have
\begin{eqnarray*}
\dot{\theta}_{v}(t^{\ast}) &=& -\sum_{u \in N(v)}{K_{uv}\sin{(\overline{\theta}_{v}-\theta_{u}(t^{\ast}))}} + \omega_{v} \\
&<& -\sum_{u \in N(v)}{\min{\left\{K_{uv}\sin{(\overline{\theta}_{v}-\theta_{u})} : \right.}} \\
&& \quad \left.|\theta_{u}| \leq \overline{\theta}_{u}\right\} + |\omega_{v}| \\
&<& -(|\omega_{v}| + \epsilon) + |\omega_{v}| < -\epsilon
\end{eqnarray*}
implying that $|\theta_{v}(t)| \leq \overline{\theta}_{v}$ in some neighborhood of $t^{\ast}$ and contradicting the definition of $t^{\ast}$.

A similar argument holds when $\theta_{v}(t^{\ast}) = -\overline{\theta}_{v}$.  We have
\begin{IEEEeqnarray*}{rCl}
\dot{\theta}_{v}(t^{\ast}) &=& -\sum_{u \in N(v)}{K_{uv}\sin{(-\overline{\theta}_{v}-\theta_{u}(t^{\ast}))}} + \omega_{v} \\
&=&\sum_{u \in N(v)}{K_{uv}\sin{(\overline{\theta}_{v}+\theta_{u}(t^{\ast}))}} + \omega_{v} \\
&>& \sum_{u \in N(v)}{\min{\left\{K_{uv}\sin{(\overline{\theta}_{v}-\theta_{u})} : \right.}} \\
&& \left. |\theta_{u}| \leq \overline{\theta}_{u}\right\} + \omega_{v} \\
&>& |\omega_{v}| + \epsilon - |\omega_{v}| > \epsilon
\end{IEEEeqnarray*}
which implies that $\theta_{v}(t) > -\overline{\theta}_{v}$ in some neighborhood of $t^{\ast}$.

These contradictions imply that $\Lambda_{PI}$ is positive invariant.
\end{IEEEproof}

Proposition \ref{prop:PI_sufficient_condition} can be interpreted as, when any oscillator $v$ reaches the boundary of $\Lambda_{PI}$, the coupling with the neighboring oscillators is sufficient to draw $v$ back to the interior of $\Lambda_{PI}$.
The proposition leads to the following procedure for computing a positive invariant set $\Lambda_{PI}$, if such a set exists, for a given set of input nodes $A$.  %In the procedure, we assume that the sets $\Lambda_{init}$ and $\Lambda_{bound}$ are defined by $\Lambda_{init} = \{\boldsymbol{\theta} : |\theta_{v}| \leq \theta_{v}^{0}\}$ and $\Lambda_{bound} = \{\boldsymbol{\theta} : |\theta_{v}| \leq \theta_{v}^{max}\}$ (if this is not the case, then we can select a superset of $\Lambda_{init}$ and a subset of $\Lambda_{bound}$ with this structure).
Select an integer $M > 0$, and define $\mathbf{X}[k]$ to be a vector in $\{0,\ldots,M\}^{n}$, where $k$ is a discrete time index.  Let $r(v) = \lceil \frac{M\theta_{v}^{0}}{\theta_{v}^{max}} \rceil$, and initialize $\mathbf{X}[0]$ as $X_{v}[0] = r(v)$ for $v \notin A$ and $X_{v}[0] = 0$ for $v \in A$.  At the $k$-th iteration of the algorithm, select $v \in V$ such that
\begin{multline}
\label{eq:sufficient_PI_algo_step}
\sum_{u \in N(v)}{\min{\left\{K_{uv}\sin{\left(\frac{\theta_{v}^{max}X_{v}[k-1]}{M} - \theta_{u}\right)} :\right.}} \\
\left. |\theta_{u}| \leq \frac{\theta_{u}^{max}X_{u}[k-1]}{M}\right\} \leq |\omega_{v}| + \epsilon.
\end{multline}
Set $X_{v}[k] = X_{v}[k-1]+1$, and $X_{u}[k] = X_{u}[k-1]$ for $u \neq v$.  The algorithm terminates when $X_{v}[k] = M$ for some $v \in V$, or when no index $v$ satisfies the conditions of (\ref{eq:sufficient_PI_algo_step}).    A pseudocode description is given as Algorithm \ref{algo:sufficient_PI}.

 \begin{center}
\begin{algorithm}[!htp]
	\caption{Algorithm for computing a positive invariant set $\Lambda_{PI}$ associated with input nodes $A$.}
	\label{algo:sufficient_PI}
	\begin{algorithmic}[1]
		\Procedure{Identify\_PI\_Set}{$G=(V,E)$, $\omega_{1},\ldots,\omega_{n}$, $A$, $\theta_{v}^{0} : v \in V$, $\theta_{v}^{max} : v\in V$}
            \State \textbf{Input}: Graph $G=(V,E)$, intrinsic frequencies $\omega_{1},\ldots,\omega_{n}$, input nodes $A$, initial states $\{\theta_{v}^{0} : v \in V\}$, boundary values $\{\theta_{v}^{max} : v \in V\}$
            \State \textbf{Output}: $(\overline{\theta}_{1},\ldots,\overline{\theta}_{n})$ such that $\{\boldsymbol{\theta} : |\theta_{v}| \leq \overline{\theta}_{v}\}$ is positive invariant, or $\emptyset$ if no such set can be found
            \State \textbf{Initialization:} $r(v) \leftarrow \lceil \frac{\theta_{v}^{0}M}{\theta_{v}^{max}}\rceil$, $X_{v}[k] \leftarrow 0$ for $v \in A$, $X_{v}[k] \leftarrow r(v)$ for $v \notin A$, $k \leftarrow 0$
           % \State $RoundKey \leftarrow Key$
           \While{$1$}
           \State $k \leftarrow k+1$, $\mathbf{X}[k] \leftarrow \mathbf{X}[k-1]$
           \For{$v \notin A$}
           \State $\rho_{uv} \leftarrow \min{\left\{\sin{\left(\frac{\theta_{v}^{max}X_{v}[k-1]}{M} - \theta_{u}\right)}: \right.}$ $\left.|\theta_{u}| \leq \frac{\theta_{u}^{max}X_{u}[k-1]}{M}\right\}$
           \State $\rho_{v} \leftarrow \sum_{u \in N(v)}{K_{uv}\rho_{uv}}$
           %$\rho_{v}$ $\leftarrow \sum_{u \in N(v)}{\min{\left\{K_{uv}\right.}}$ \sin{\left(\frac{\theta_{v}^{max}X_{v}[k-1]}{M} - \theta_{u}\right)}\right.}}  :$ $\left.|\theta_{u}| \leq \frac{\theta_{u}^{max}X_{u}[k-1]}{M}\right\}$
           \EndFor
           \If{$\rho_{v} > |\omega_{v}| + \epsilon \ \forall v$}
           %\For{$v \in V$}
           %\State $\overline{\theta}_{v} \leftarrow \frac{\theta_{v}^{max}}{M}X_{i}[k-1]$
           \State \Return{$\left(\frac{\theta_{1}^{max}X_{1}[k-1]}{M},\ldots,\frac{\theta_{n}^{max}X_{n}[k-1]}{M}\right)$}
           %\EndFor
           \Else
           \State Choose $v$ such that $\rho_{v} \leq |\omega_{v}| + \epsilon$
           \State $X_{v}[k] \leftarrow X_{v}[k-1]+1$%, $X_{j}[k] \leftarrow X_{j}[k-1]$ for $j \neq i$
           \IIf{$X_{v}[k]==M$} \Return{$\emptyset$} \EndIIf
           %\EndIf
           \EndIf
           \EndWhile
		\EndProcedure
	 \end{algorithmic}
\end{algorithm}
\end{center}

The algorithm discretizes the interval $[0,\gamma]$ by dividing it into $M$ intervals.  Hence, a larger value of $M$ results in more granular intervals and a more precise approximation of the positive invariant set.

The following theorem describes the guarantees provided by Algorithm \ref{algo:sufficient_PI}.
\begin{theorem}
\label{theorem:sufficient_PI_algorithm}
Suppose that Algorithm \ref{algo:sufficient_PI} returns a nonempty vector  $(\overline{\theta}_{1},\ldots,\overline{\theta}_{n})$.  The set $\Lambda_{PI} = \{\boldsymbol{\theta} : |\theta_{v}| \leq \overline{\theta}_{v}\}$ is positive invariant and satisfies $\Lambda_{init} \subseteq \Lambda_{PI} \subseteq \Lambda_{bound}$.
\end{theorem}

\begin{IEEEproof}
The approach is to show that $\Lambda_{PI}$ satisfies the conditions of Proposition \ref{prop:PI_sufficient_condition}.
Suppose that there exists $v$ such that $(\ref{eq:PI_condition})$ does not hold.  Then $$\sum_{u \in N(v)}{\min{\left\{K_{uv}\sin{\left(\overline{\theta}_{v}-\theta_{u}\right)} : |\theta_{u}| \leq \overline{\theta}_{u}\right\}}} \leq |\omega_{v}| + \epsilon,$$ and hence Algorithm \ref{algo:sufficient_PI} would increment the value of $X_{v}[k]$ instead of returning $(\overline{\theta}_{1},\ldots,\overline{\theta}_{n})$.  This contradiction implies that $\Lambda_{PI}$ is positive invariant.

We now show that $\Lambda_{init} \subseteq \Lambda_{PI} \subseteq \Lambda_{bound}$.  For each $v \notin A$, $$\overline{\theta}_{v} = \frac{\theta_{v}^{max}X_{v}[k]}{M} \geq \frac{\theta_{v}^{max}r(v)}{M} \geq \theta_{v}^{0},$$ implying that $\Lambda_{init} \subseteq \Lambda_{PI}$.  Furthermore, if Algorithm \ref{algo:sufficient_PI} does not return $\emptyset$, then $X_{v}[k] < M$ when the algorithm terminates and hence $\overline{\theta}_{v} < \theta_{v}^{max}$, implying that $\Lambda_{PI} \subseteq \Lambda_{bound}$.
\end{IEEEproof}

Algorithm \ref{algo:sufficient_PI} provides an efficient procedure for verifying existence of a positive invariant set $\Lambda_{PI}$ and constructing such a set for a given set of input nodes $A$.  In addition, this procedure will form the basis for the submodular optimization approach to selecting input nodes presented in Section \ref{subsec:submod_PI}.

\subsection{Verifying Convergence to $\Lambda_{final}$}
\label{subsec:sufficient_convergence}
The following proposition gives sufficient conditions to ensure that, for any $\boldsymbol{\theta}(0) \in \Lambda_{init}$, there exists $T$ such that $\boldsymbol{\theta}(t) \in \Lambda_{final}$ for all $t \geq T$.  %We then present an algorithm for verifying this sufficient for a given input set.

\begin{proposition}
\label{prop:sufficient_convergence}
Let $\Lambda = \{\boldsymbol{\theta} : |\theta_{v}| \leq \overline{\theta}_{v}\}$ be a positive invariant set.  Suppose there exists an index $v$ and $\underline{\theta}_{v} \in [0, \overline{\theta}_{v}]$ such that
\begin{multline}
\label{eq:sufficient_convergence}
\sum_{u \in N(v)}{\min{\left\{K_{uv}\sin{(\theta_{v}-\theta_{u})}\right.}} : \\
 \left.\theta_{v} \in [\underline{\theta}_{v},\overline{\theta}_{v}], |\theta_{u}| \leq \overline{\theta}_{u}\right\} > |\omega_{v}| + \epsilon.
 \end{multline}
 Define $\Lambda^{\prime} = \Lambda \setminus \{\boldsymbol{\theta} : \theta_{v} \in [\underline{\theta}_{v}, \overline{\theta}_{v}]\}$.  Then $\Lambda^{\prime}$ is positive invariant and, when $\boldsymbol{\theta}(0) \in \Lambda$, there exists $T$ such that $\boldsymbol{\theta}(t) \in \Lambda^{\prime}$ for all $t \geq T$.
\end{proposition}

\begin{IEEEproof}
We first show positive invariance of $\Lambda^{\prime}$.  Suppose $\boldsymbol{\theta}(0) \in \Lambda^{\prime}$. Since $\Lambda$ is positive invariant, $|\theta_{u}| \leq \overline{\theta}_{u}$ for all $u \neq v$.  It suffices to show that if $|\theta_{v}(0)| \leq \underline{\theta}_{v}$, then $|\theta_{v}(t)| \leq \underline{\theta}_{v}$ for all $t \geq 0$.
%Since the set $\Lambda$ is assumed to be positive invariant, in order to prove positive invariance of $\Lambda^{\prime}$, it suffices to show that if $\boldsymbol{\theta}(0) \in \Lambda$ and $|\theta_{v}(0)| \leq \underline{\theta}_{v}$, then $\theta_{v}(t) \leq \underline{\theta}_{v}$ for all $t \geq 0$.
 This result holds by (\ref{eq:sufficient_convergence}) and Proposition \ref{prop:PI_sufficient_condition}.

It remains to prove that, if $\boldsymbol{\theta}(0) \in \Lambda$, then there exists $T$ such that $\boldsymbol{\theta}(t) \in \Lambda^{\prime}$ for $t \geq T$.  Suppose that this is not the case, and choose $\boldsymbol{\theta}(0) \in \Lambda$ such that $\theta_{v}(t) \in [\underline{\theta}_{v}, \overline{\theta}_{v}]$ for all $t \geq 0$.  Then
\begin{IEEEeqnarray*}{rCl}
\dot{\theta}_{v}(t) &=& -\sum_{u \in N(v)}{K_{uv}\sin{(\theta_{v}(t)-\theta_{u}(t))}} + \omega_{v} \\
&\leq& -|\omega_{v}|-\epsilon + |\omega_{v}| = -\epsilon,
\end{IEEEeqnarray*}
implying that $$\theta_{v}(t) \leq \theta_{v}(0) - \epsilon t \leq \overline{\theta}_{v} - \epsilon t$$ and hence $\theta_{v}(t) \leq \underline{\theta}_{v}$ for $t$ sufficiently large.  This contradiction implies that $\boldsymbol{\theta}(T) \in \Lambda^{\prime}$ for some $T > 0$.  Positive invariance of $\Lambda^{\prime}$ then yields $\boldsymbol{\theta}(t) \in \Lambda^{\prime}$ for $t \geq T$.
\end{IEEEproof}

We now present an efficient algorithm for checking whether, for any $\boldsymbol{\theta}(0) \in \Lambda_{init}$, there exists $T \geq 0$ such that $\boldsymbol{\theta}(t) \in \Lambda_{final}$ for all $t \geq T$.   We assume that Algorithm \ref{algo:sufficient_PI} has already been used to construct a set $\Lambda_{PI}$, satisfying the conditions (a) and (b) of Theorem \ref{theorem:sufficient_practical_sync}, of the form $\Lambda_{PI} = \{\boldsymbol{\theta} : |\theta_{v}| \leq \theta_{v}^{PI}\}$.

Let $M$ be defined as in Section \ref{subsec:conditions_PI}.  For each $v$, let $r(v) = \lfloor \frac{M\gamma}{\theta_{v}^{max}}\rfloor$ and $s(v) = \lceil \frac{M\theta_{v}^{PI}}{\theta_{v}^{max}}\rceil$.  Define a vector $\mathbf{X}[k] \in \{0,\ldots,M\}^{n}$, where $k$ is a time index.  Define $X_{v}[0] = s(v)$ for $v \notin A$ and $X_{v}[0] = 0$ for $v \in A$.  At time step $k$, select an index $v$ such that
\begin{multline}
\label{eq:sufficient_convergence_condition}
\sum_{u \in N(v)}{\min{\left\{K_{uv}\sin{(\theta_{v}-\theta_{u})}:\right.}} \\
  \theta_{v} \in \left[\frac{(X_{v}[k-1]-1)\theta_{v}^{max}}{M}, \frac{X_{v}[k-1]\theta_{v}^{max}}{M}\right],\\
 \left. |\theta_{u}| \leq \frac{X_{u}[k-1]\theta_{u}^{max}}{M}\right\}
  > |\omega_{v}| + \epsilon
\end{multline}
 if such an index exists.  Set $X_{v}[k] = X_{v}[k-1]-1$ and $X_{u}[k] = X_{u}[k-1]$ if $u \neq v$.  The algorithm terminates if no index $v$ satisfying (\ref{eq:sufficient_convergence_condition}) can be found.  A pseudocode description of the algorithm is given as Algorithm \ref{algo:sufficient_convergence}.

  \begin{center}
\begin{algorithm}[!htp]
	\caption{Algorithm for proving that a set of input nodes $A$ is sufficient to ensure convergence to $\Lambda_{final}$.}
	\label{algo:sufficient_convergence}
	\begin{algorithmic}[1]
		\Procedure{Check\_Convergence}{$G=(V,E)$, $\omega_{1},\ldots,\omega_{n}$, $A$, $\theta_{v}^{0} : v \in V$, $\theta_{v}^{max} : v \in V$, $\theta_{v}^{\prime}: v \in V$, $\theta_{v}^{PI} : v \in V$}
            \State \textbf{Input}: Graph $G=(V,E)$, intrinsic frequencies $\omega_{1},\ldots,\omega_{n}$, input nodes $A$, initial states $\{\theta_{v}^{0} : v \in V\}$, boundary values $\{\theta_{v}^{max} : v \in V\}$, positive invariant set defined by $\{\theta_{v}^{PI} : v \in V\}$, set $\Lambda_{final}$ defined by $\{\theta_{v}^{\prime} : v \in V\}$.
            \State \textbf{Output}: Return \textbf{true} if, for any $\boldsymbol{\theta} \in \Lambda_{PI}$, $\boldsymbol{\theta}(t) \in \Lambda_{final}$ for $t$ sufficiently large.  Return \textbf{false} otherwise. %$(\overline{\theta}_{1},\ldots,\overline{\theta}_{n})$ such that, for any $\boldsymbol{\theta} \in \Lambda_{init}$, $|\theta_{v}(t)| \leq \overline{\theta}_{v}$ for $t$ sufficiently large
            \State \textbf{Initialization:} $r(v) \leftarrow \lfloor \frac{M\theta_{v}^{\prime}}{\theta_{v}^{max}}\rfloor$, $s(v) \leftarrow \lceil \frac{M\theta_{v}^{PI}}{\theta_{v}^{max}}\rceil$, $k \leftarrow 0$, $X_{v}[k] \leftarrow 0$ for $v \in A$, $X_{v}[k] \leftarrow s(v)$ for $v \notin A$
           % \State $RoundKey \leftarrow Key$
           \While{$1$}
           \State $k \leftarrow k+1$, $\mathbf{X}[k] \leftarrow \mathbf{X}[k-1]$
           \For{$v \notin A$}
           \State $\underline{\theta}_{v} \leftarrow \frac{(X_{v}[k-1]-1)\theta_{v}^{max}}{M}$
           \State $\overline{\theta}_{v} \leftarrow \frac{X_{v}[k-1]\theta_{v}^{max}}{M}$
           \For{$u \in N(v)$}
           \State $\overline{\theta}_{u} \leftarrow \frac{X_{u}[k-1]\theta_{u}^{max}}{M}$
           \State $\rho_{uv} \leftarrow \min{\left\{\sin{(\theta_{v}-\theta_{u})} :  \theta_{v} \in [\underline{\theta}_{v}, \overline{\theta}_{v}], |\theta_{u}| \leq \overline{\theta}_{u}\right\}}$
           \EndFor
           \State $\rho_{v} \leftarrow \sum_{u \in N(v)}{K_{uv}\rho_{uv}}$
           \EndFor
           \If{$\rho_{v} \leq |\omega_{v}| + \epsilon \ \forall v$}
           %\For{$i=1,\ldots,n$}
           %\State $\overline{\theta}_{v} \leftarrow \frac{\theta_{v}^{max}}{M}X_{i}[k-1]$
           \State $\overline{\theta}_{v} \leftarrow \frac{\theta_{v}^{max}X_{v}[k-1]}{M} \ \forall v \in V$
           \If{$\overline{\theta}_{v} \leq \theta_{v}^{\prime}$ for all $v \in V$}
           \State \Return{\textbf{true}}
           \Else
           \State \Return{\textbf{false}}
           \EndIf
           %\EndFor
           \Else
           \State Choose $v$ such that $\rho_{v} > |\omega_{v}| + \epsilon$
           \State $X_{v}[k] \leftarrow X_{v}[k-1]-1$%, $X_{j}[k] \leftarrow X_{j}[k-1]$ for $j \neq i$
           %\IIf{$X_{i}[k]==M$} \Return{$\emptyset$} \EndIIf
           %\EndIf
           \EndIf
           \EndWhile
		\EndProcedure
	 \end{algorithmic}
\end{algorithm}
\end{center}

The following theorem describes how the results of Algorithm \ref{algo:sufficient_convergence} can be used to verify convergence to $\Lambda_{final}$.

\begin{theorem}
\label{theorem:algo_sufficient_convergence}
Suppose that there exists a positive invariant set $\Lambda_{PI}$, and that $\overline{\boldsymbol{\theta}} = (\overline{\theta}_{1},\ldots,\overline{\theta}_{n})$ is returned by Algorithm \ref{algo:sufficient_convergence}.  Let $\overline{\Lambda} = \{\boldsymbol{\theta} : |\theta_{v}| \leq \overline{\theta}_{v}\}$.  If $\overline{\Lambda} \subseteq \Lambda_{final}$, then for each $\boldsymbol{\theta}(0) \in \Lambda_{init}$, there exists $T$ such that $\boldsymbol{\theta}(t) \in \Lambda_{final}$ for all $t \geq T$.
\end{theorem}

\begin{IEEEproof}
Let $\Lambda_{k} = \left\{\boldsymbol{\theta} : |\theta_{v}| \leq \frac{X[k]\theta_{v}^{max}}{M}\right\}$.  The approach of the proof is to show that for all $k$, $\boldsymbol{\theta}(0) \in \Lambda_{PI}$ implies that there exists $T$ with $\boldsymbol{\theta}(t) \in \Lambda_{k}$ for $t \geq T$.

The proof is by induction on $k$.  When $k=0$, $\Lambda_{0} = \Lambda_{PI}$, and hence the result holds by the positive invariance of $\Lambda_{PI}$.  At step $k$, the index $v$ selected by the algorithm satisfies (\ref{eq:sufficient_convergence_condition}), and hence Proposition \ref{prop:sufficient_convergence} holds with $\Lambda = \Lambda_{k-1}$, $\overline{\theta}_{v} = \frac{X_{v}[k-1]\theta_{v}^{max}}{M}$, and $\underline{\theta}_{v} = \frac{(X_{v}[k-1]-1)\theta_{v}^{max}}{M}$.  In this case, $\Lambda_{k-1} \setminus \{\boldsymbol{\theta} : \theta_{v} \in [\underline{\theta}_{v}, \overline{\theta}_{v}]\} = \Lambda_{k}$.  By Proposition \ref{prop:sufficient_convergence}, $\boldsymbol{\theta}(t) \in \Lambda_{k}$ for $t$ sufficiently large, establishing the inductive step.

The fact that $\boldsymbol{\theta}(t) \in \Lambda_{final}$ for $t$ sufficiently large then follows from $\overline{\Lambda} \subseteq \Lambda_{final}$.
\end{IEEEproof}

If Algorithm \ref{algo:sufficient_convergence} returns \textbf{False}, then the set of input nodes $A$ is insufficient to guarantee convergence to $\Lambda_{final}$, and hence a new input set must be selected.  Selecting such a set is the topic of Section \ref{sec:submodular}.

\subsection{Example}
\label{subsec:example_verification}
Consider a line network, with $V = \{n_{1},n_{2},n_{3}\}$ and $E = \{(n_{1}, n_{2}), (n_{2}, n_{3})\}$, in which the set of input nodes is given by $A = \{n_{1}\}$.  In the network, $K_{uv} = 1$ for all $(i,j) \in E$, and set $\omega_{1} = 0.0705$, $\omega_{2} = 0.0709$, and $\omega_{3} = 0.0336$.  The set $\Lambda_{bound} = \{||\boldsymbol{\theta}||_{\infty} \leq \frac{\pi}{4}\}$, $\Lambda_{init} = \{||\boldsymbol{\theta}||_{\infty} \leq \frac{\pi}{5}\}$, and $\Lambda_{final} = \{||\boldsymbol{\theta}||_{\infty} \leq \frac{\pi}{6}\}$.

Set $M=20$ and consider Algorithm \ref{algo:sufficient_PI}.  With $M=20$, $\mathbf{X}[0] = (0, 16, 16)$.  At the first iteration, $X_{3}$ is incremented, resulting in $\mathbf{X}[1] = (0,16,17)$.  The algorithm terminates, and hence $\Lambda_{PI}$ is defined by $\overline{\boldsymbol{\theta}} = \{0, \frac{\pi}{5}, \frac{17\pi}{80}\}$.

\begin{figure}[!ht]
\centering
\includegraphics[width=3in]{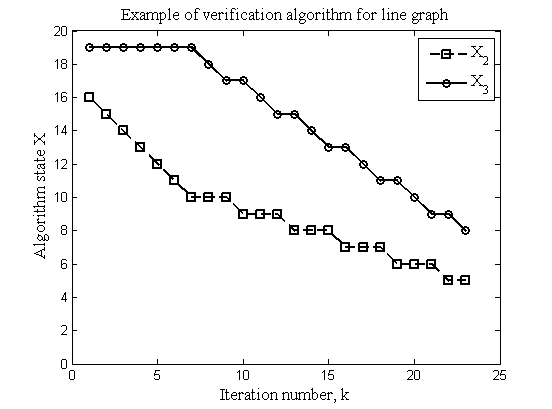}
\caption{Values of $X_{2}$ and $X_{3}$ at each iteration of Algorithm \ref{algo:sufficient_convergence} for a line graph ($n_{1}$ is the input and hence $X_{1}[k]$ is identically zero).}
\label{fig:sufficient_verify_example}
\end{figure}

Applying Algorithm \ref{algo:sufficient_convergence} with these values of $\Lambda_{PI}$ proceeds as follows.  This time, $\mathbf{X}[0] = (0,16,17)$. The values of $\mathbf{X}$ at subsequent iterations are shown in Figure \ref{fig:sufficient_verify_example}.

The values of $X_{2}$ and $X_{3}$ converge to $5$ and $8$, respectively, while the upper bound established by $\Lambda_{final}$ is $13$. We observe that the values of $X_{2}$ and $X_{3}$ alternate, with $X_{2}$ being decremented, followed by $X_{3}$, followed by $X_{2}$, and so on until convergence occurs.  This demonstrates the impact of the synchronization of a node's neighbors on the synchronization of that node.  Also, observe that $X_{3} > X_{2}$.  This is because the input node $n_{1}$ is directly connected to node $n_{2}$, and hence node $n_{2}$ is closer to synchronization than node $n_{3}$.

This section developed sufficient conditions for practical synchronization with a given set of input nodes.  In the following section, we will propose algorithms for selecting input nodes to guarantee practical synchronization based on these conditions.

\section{Submodular Optimization Approach to Input Selection}
\label{sec:submodular}
In this section, we present our submodular optimization approach to selecting a set of input nodes to ensure practical synchronization.  Our approach is based on ensuring that the conditions for existence of a positive invariant set (Section \ref{subsec:conditions_PI}) and convergence of the node phases to $\Lambda_{final}$ (Section \ref{subsec:sufficient_convergence}) are satisfied by establishing an equivalence between both conditions and the connectivity of an augmented network graph.  %guarantee practical node or edge synchronization from almost any initial state.  Our approach is based on ensuring that the conditions for existence of a cohesive equilibrium (Section \ref{subsec:verify_existence}) and non-existence of stable non-cohesive equilibria (Section \ref{subsec:non_cohesive_algo}) are satisfied.

Section \ref{subsec:submod_PI} formulates the condition for existence of a positive invariant set $\Lambda_{PI}$ that satisfies $\Lambda_{init} \subseteq \Lambda_{PI} \subseteq \Lambda_{bound}$ as a submodular constraint.  We first define an augmented graph and prove that the conditions for existence of a positive invariant set in Theorem \ref{theorem:sufficient_PI_algorithm} are equivalent to the connectivity of a class of subgraphs of the augmented graph.  We then show that this connectivity condition can be expressed as a constraint of the form $h_{1}(A) = 0$, where $h_{1}$ is a supermodular function of $A$.  We recall that a function $h: 2^{V} \rightarrow \mathbb{R}$ is \emph{submodular} if, for any sets $A$ and $B$ with $A \subseteq B$ and any $v \notin B$, $$h(A \cup \{v\}) - h(A) \geq h(B \cup \{v\}) - h(B)$$ and that a function $h$ is \emph{supermodular} if $-h$ is submodular.

Section \ref{subsec:submod_convergence} formulates convergence to the set $\Lambda_{final}$ as a submodular constraint on the set of input nodes.  As in Section \ref{subsec:submod_PI}, we first construct an augmented graph, and prove that the conditions identified in Theorem \ref{theorem:algo_sufficient_convergence} are equivalent to connectivity of a class of subgraphs of the augmented graph (albeit a different class from that identified in Section \ref{subsec:submod_PI}).  We show that this connectivity condition is equivalent to a submodular constraint on the input set.

%Section \ref{subsec:submod_existence} formulates the condition for existence of a cohesive equilibrium as a submodular constraint on the input set.  We first define an augmented graph and prove that the conditions for existence shown in Section \ref{subsec:verify_existence} are equivalent to connectivity of a class of random subgraphs of the augmented graph.  We then show that this connectivity condition can be expressed as a constraint of the form $h(A) = n$, where $h$ is a submodular function.  We recall that a function $h: 2^{V} \rightarrow \mathbb{R}$ is \emph{submodular} if, for any sets $A$ and $B$ with $A \subseteq B$ and any $v \notin B$, $$h(A \cup \{v\})- h(A) \geq h(B \cup \{v\}) - h(B).$$

%Section \ref{subsec:submod_non_existence} formulates non-existence of stable non-cohesive equilibria as a submodular constraint on the input set.  As in Section \ref{subsec:submod_existence}, we first construct an augmented graph, although the construction of this section differs from Section \ref{subsec:submod_non_existence}.   We then prove that the conditions identified in Section \ref{subsec:non_cohesive_algo} are equivalent to connectivity of a class of subgraphs of the augmented graph, and show that this connectivity condition is equivalent to a submodular constraint on the input set.

In Section \ref{subsec:input_selection}, we formulate the problem of selecting a set of input nodes to achieve practical synchronization as an optimization problem that combines the constraints of Sections \ref{subsec:submod_PI} and \ref{subsec:submod_convergence}.  We then present algorithms for selecting a set of input nodes and analyze the optimality guarantees of those algorithms.

\subsection{Submodular Constraint on Input Nodes for Positive Invariance}
\label{subsec:submod_PI}
We first define a weighted augmented network graph, denoted $\tilde{G} = (\tilde{V},\tilde{E})$, and then formulate conditions for existence of a positive invariant set $\Lambda_{PI}$ based on this graph.  Let $M > 0$ be a positive integer as in Section \ref{sec:conditions}.  The vertex set of $\tilde{G}$ consists of $M$ copies of the vertices of $G$, i.e., $\tilde{V} = \{v_{m} : m=0,\ldots,M, v \in V\}$.  The edge set $\tilde{E}$ is defined by
\begin{multline*}
\tilde{E} = \{(u_{l},v_{m}) : (u,v) \in E, l=0,\ldots,M, m=0,\ldots,M\}.
\end{multline*}
The graph $\tilde{G}$ is directed, with edge $(u_{l},v_{m}) \in \tilde{E}$ implying a directed edge from $u_{l}$ to $v_{m}$.
Each edge $(u_{l},v_{m}) \in \tilde{E}$ has an associated weight, defined as follows.  As a preliminary, let
\begin{multline*}
\tilde{\alpha}_{uv}(m,l) \\
 \triangleq K_{uv}\left(1- \min{\left\{\sin{\left(\frac{\theta_{v}^{max}m}{M}-\theta_{u}\right)} : |\theta_{u}| \leq \frac{\theta_{u}^{max}l}{M}\right\}}\right).
\end{multline*}
For each edge $(u_{l},v_{m})$, define the weight $\tilde{\beta}_{uv}(m,l) = \tilde{\alpha}_{uv}(m,l+1)-\tilde{\alpha}_{uv}(m,l)$.  When $l=M$, define $\tilde{\beta}_{uv}(m,M) = 0$.  Note that the weights $\tilde{\beta}_{uv}(m,l)$ are nonnegative.

A class of subgraphs of $\tilde{G}$ is defined as follows.

\begin{definition}
\label{def:class_T_definition}
Let $G^{\prime}$ be a subgraph of $\tilde{G}$. The subgraph $G^{\prime}$ is a \emph{class}-$\mathcal{T}$ subgraph if the neighbor set $N^{\prime}(v_{m})$ of each node $v_{m}$ with $v \notin A$ satisfies
\begin{equation}
\label{eq:class_T_definition}
\sum_{u_{l} \in N^{\prime}(v_{m})}{\tilde{\beta}_{uv}(m,l)} > \sum_{u \in N(v)}{\sum_{l=0}^{M-1}{\tilde{\beta}_{uv}(m,l)}} - \tau_{v}^{m},
\end{equation}
where $$\tau_{v}^{m} \triangleq \sum_{u \in N(v)}{K_{uv}} - |\omega_{v}| - \epsilon-\sum_{u \in N(v)}{\alpha_{uv}(m,0)}.$$ If no such subset exists, then $N^{\prime}(v_{m}) = \tilde{N}(v_{m})$. For each node $v \in A$, the neighbor set $N^{\prime}(v_{m}) = \emptyset$.
\end{definition}

If there exists a node $v$ such that no neighbor set $N^{\prime}(v_{m})$ satisfies (\ref{eq:class_T_definition}) for all $m=0,ldots,M$, then node $v$ must be chosen as an input to ensure practical synchronization.  The following theorem gives conditions for convergence based on Definition \ref{def:class_T_definition}.

\begin{theorem}
\label{theorem:PI_aug_graph}
Define $r(v) = \lceil \frac{M\theta_{v}^{0}}{\theta_{v}^{max}} \rceil$, $V^{\prime} = \{v_{r(v)} : v \in V\}$, and $V^{\prime\prime} = \{v_{M} : v \in V\}$.
If there exists a class-$\mathcal{T}$ subgraph in which $v_{r(v)}$ is not connected to $V^{\prime\prime}$ for all $v \in V$, then there exists a positive invariant set $\Lambda_{PI}$ satisfying $\Lambda_{init} \subseteq \Lambda_{PI} \subseteq \Lambda_{bound}$.
\end{theorem}

\begin{IEEEproof}
Suppose that such a class-$\mathcal{T}$ subgraph, denoted $G^{\prime}$, exists.  For each $v \in V$, define $m_{v}^{\ast} = \max{\{m : v_{m} \mbox{ connected to } v_{r(v)}\}}$, and let $\hat{m}_{v} = m_{v}^{\ast} + 1$.  In $G^{\prime}$, $N^{\prime}(v_{\hat{m}_{v}}) \cap \{u_{l} : l \leq m_{u}^{\ast}\} = \emptyset$ (otherwise, $v_{\hat{m}_{v}}$ would be connected to $v_{r(v)}$, a contradiction).  Since the sets are disjoint and their union at most covers $N(v_{m})$, the sum of the nonnegative weights of the edges in $N^{\prime}(v_{r(v)})$ and the edges in $\{u_{l} : l \leq m_{u}^{\ast}\}$ is upper bounded by the sum of the edge weights in $N(v_{r(v)})$.  Hence
\begin{multline*}
\sum_{u_{l} \in N^{\prime}(v_{\hat{m}_{v}})}{\beta_{uv}(\hat{m}_{v},l)} + \sum_{u \in N(v)}{\sum_{l=0}^{m_{u}^{\ast}}{\tilde{\beta}_{uv}(\hat{m}_{v},l)}} \\
\leq \sum_{u \in N(v)}{\sum_{l=0}^{M-1}{\tilde{\beta}_{uv}(m,l)}}.
\end{multline*}
Substituting (\ref{eq:class_T_definition}) into the first term of the left-hand side yields
\begin{multline*}
\sum_{u \in N(v)}{\sum_{l=0}^{M-1}{\tilde{\beta}_{uv}(m,l)}} - \tau_{v}^{\hat{m}_{v}} + \sum_{u \in N(v)}{\sum_{l=0}^{m_{u}^{\ast}}{\tilde{\beta}_{uv}(\hat{m}_{v},l)}}\\
 < \sum_{u \in N(v)}{\sum_{l=0}^{M-1}{\tilde{\beta}_{uv}(m,l)}}.
 \end{multline*}
 Rearranging terms and substituting the definition of $\tau_{v}^{\hat{m}_{v}}$, we have $$\sum_{u \in N(v)}{\sum_{l=0}^{m_{u}^{\ast}}{\tilde{\beta}_{uv}(\hat{m}_{v},l)}} < \sum_{u \in N(v)}{K_{uv}} - |\omega_{v}| - \sum_{u \in N(v)}{\alpha_{uv}(\hat{m}_{v},0)}$$ which is equivalent to
 \begin{IEEEeqnarray*}{rCl}
 \sum_{u \in N(v)}{\alpha_{uv}(\hat{m}_{v}, m_{u}^{\ast}+1)} &=& \sum_{u \in N(v)}{\alpha_{uv}(\hat{m}_{v}, \hat{m}_{u})} \\
  &>& \sum_{u \in N(v)}{K_{uv}} - |\omega_{v}| - \epsilon.
 \end{IEEEeqnarray*}
   Substituting the definition of $\alpha_{uv}(\hat{m}_{v}, \hat{m}_{u})$ and subtracting $\sum_{u \in N(v)}{K_{uv}}$ from both sides yields
\begin{multline*}
\sum_{u \in N(v)}{K_{uv}\min{\left\{\sin{(\frac{\hat{m}_{v}\theta_{v}^{max}}{M} - \theta_{u})} : |\theta_{j}| \leq \frac{\hat{m}_{u}\theta_{u}^{max}}{M}\right\}}} \\
 > |\omega_{v}| + \epsilon.
 \end{multline*}
   Hence the set $$\Lambda_{PI} = \left\{\boldsymbol{\theta} : |\theta_{v}| \leq \frac{\hat{m}_{v}\theta_{v}^{max}}{M}\right\}$$ satisfies the conditions of Proposition \ref{prop:PI_sufficient_condition}, and is  positive invariant.
\end{IEEEproof}

We now formulate sufficient conditions for a set of input nodes to guarantee the existence of a positive invariant set, based on Theorem \ref{theorem:PI_aug_graph}.  For a class-$\mathcal{T}$ subgraph, the impact of adding a node $v$ to the set $A$ will be to remove the set of edges $(u_{l},v_{m})$ for all $u_{l} \in N^{\prime}(v_{m})$ and $m=0,\ldots,M$.  Adding enough nodes to the input set will create a cut in the graph, disconnecting $V^{\prime}$ and $V^{\prime\prime}$.  Formally, this intuition is described by the following lemma.

\begin{lemma}
\label{lemma:PI_graph_cut}
  If the nodes $\{v_{m}: v \in A, m=0,\ldots,M\}$ form a cut between $V^{\prime}$ and $V^{\prime\prime}$ in any class-$\mathcal{T}$ subgraph $G^{\prime}$ of $\tilde{G}$, then there exists a positive-invariant set $\Lambda_{PI}$ satisfying $\Lambda_{init} \subseteq \Lambda_{PI} \subseteq \Lambda_{bound}$.
\end{lemma}

\begin{IEEEproof}
By Definition \ref{def:class_T_definition}, the nodes in the set $\{v_{m} : m=0,\ldots,M\}$ have $N^{\prime}(v_{m}) = \emptyset$ in any class-$\mathcal{T}$ subgraph.  Hence, if these nodes form a cut between $V^{\prime}$ and $V^{\prime\prime}$, then there will be no path from any node in $V^{\prime}$ to $V^{\prime\prime}$.  By Theorem \ref{theorem:PI_aug_graph}, there exists a positive-invariant set $\Lambda_{PI}$ satisfying $\Lambda_{init} \subseteq \Lambda_{PI} \subseteq \Lambda_{bound}$.
\end{IEEEproof}

The next step is to express the cut condition as a submodular constraint on the set of input nodes $A$.  In order to construct such a constraint, define a random walk $R[s]$ on the graph $G^{\prime}$, in which
\begin{multline*}
Pr(R[s] = v_{m} | R[s-1] = u_{l}) \\
= \left\{
\begin{array}{ll}
\frac{1}{|\{v^{\prime}_{m^{\prime}} : u_{l} \in N^{\prime}(v^{\prime}_{m^{\prime}})\}|}, & u_{l} \in N^{\prime}(v_{m}) \\
0, & \mbox{else}
\end{array}
\right.
\end{multline*}
Define $h_{1}(A,G^{\prime})$ to be the probability that a walk originating in $V^{\prime}$ reaches a node in $V^{\prime\prime}$ before arriving at any node in $\{v_{m} : v \in A, m=0,\ldots,M\}$. The function $h_{1}(A, G^{\prime})$ can be computed efficiently using methods for computing the absorption probabilities of random walks \cite{doyle2000random}. We have the following result.
%Define $\nu(G^{\prime},A)$ to be the stationary distribution of this walk, conditioned on the walk originating in $V^{\prime}$, and let $h_{1}(A, G^{\prime}) = \sum_{v_{m} \in V^{\prime\prime}}{\nu(G^{\prime},A)_{v_{m}}}$.  We have the following result.
\begin{lemma}
\label{lemma:PI_cut}
The nodes $\{v_{m} : v \in A, m=0,\ldots,M\}$ form a cut between $V^{\prime}$ and $V^{\prime\prime}$ if and only if $h_{1}(A,G^{\prime}) = 0$.
\end{lemma}

\begin{IEEEproof}
If the nodes $\{v_{m} : v \in A, m = 0,\ldots,M\}$ form a cut between $V^{\prime}$ and $V^{\prime\prime}$, then any walk will reach $v_{m}$ for some $v \in A$ before reaching $V^{\prime\prime}$, and hence $h_{1}(A, G^{\prime}) = 0$.  Conversely, if the nodes do not form a cut, then there is a sample path of the random walk that avoids all nodes in $\{v_{m} : v \in A, m = 0,\ldots,M\}$ before reaching $V^{\prime\prime}$.  This implies $h_{1}(A, G^{\prime}) > 0$.
%The stationary distribution of the random walk $R[s]$ will assign nonzero mass to each node that is path-connected to $V^{\prime}$.  Hence, the stationary distribution will be zero for all nodes in $V^{\prime\prime}$ if and only if $V^{\prime\prime}$ is disconnected from $V^{\prime}$.
\end{IEEEproof}

The following proposition leads to a submodular optimization formulation for ensuring positive invariance.

\begin{proposition}
\label{prop:PI_submodular}
  The function $h_{1}(A,G^{\prime})$ is supermodular as a function of $A$.
\end{proposition}
\begin{IEEEproof}
To show submodularity of $h_{1}(A,G^{\prime})$, consider one possible sample path of the walk.  The nodes in $\{v_{m} : v \in A\}$ are absorbing states of the walk.  Let $\chi(A)$ denote the indicator function of the event that the walk reaches any node in $V^{\prime\prime}$ before an absorbing state, so that $\chi(A) - \chi(A \cup \{v\})$ is equal to $1$ if the walk reaches a node in $V^{\prime\prime}$ before $A$, but reaches $v$ before any node in $V^{\prime\prime}$.  If $A \subseteq B$, then any walk that reaches $v$ before any node in $B$ automatically reaches $v$ before any node in $A$, and  hence $$\chi(A \cup \{v\}) - \chi(A) \leq \chi(B \cup \{v\}) - \chi(B),$$ implying that the indicator function is supermodular as a function of $A$.  Since $h_{1}$ is a nonnegative weighted sum of these indicator functions, $h_{1}(A,G^{\prime})$ is supermodular as well.
\end{IEEEproof}

Proposition \ref{prop:PI_submodular} implies that a sufficient condition for existence of a positive invariant set is $h_{1}(A,G^{\prime}) = 0$ for some class-$\mathcal{T}$ subgraph $G^{\prime}$.  Hence, one approach to ensuring the existence of a positive invariant set is to generate a collection of class-$\mathcal{T}$ subgraphs and, for each subgraph, choose a set $A$ that satisfies $h_{1}(A,G^{\prime})=0$ using submodular optimization techniques.  This approach will be formally developed in Section \ref{subsec:input_selection}.

\subsection{Submodular Constraint for Convergence to $\Lambda_{final}$}
\label{subsec:submod_convergence}
In this section, we develop a submodular optimization approach to selecting a set of input nodes $A$ that satisfies the sufficient conditions for convergence to $\Lambda_{final}$ introduced in Section \ref{subsec:sufficient_convergence}.  As in the previous section, our approach is to introduce an augmented graph $\tilde{G}$, and then provide sufficient conditions for convergence based on subgraphs of $\tilde{G}$.

Define the set $\tilde{V} = \{v_{m} : v \in V, m=0,\ldots,M\}$.  The edge set is defined by
\begin{equation*}
\tilde{E} = \{(u_{l},v_{m}) : (u,v) \in E, l=0,\ldots,M, m=0,\ldots,M\}
%\cup \{(v_{l}, v_{m}), l < m\}.
\end{equation*}
The weights of the graph $\tilde{G} = (\tilde{V}, \tilde{E})$ are defined as follows.  Define
\begin{multline*}
\alpha_{uv}(m,l) = \min{\left\{K_{uv}\sin{(\theta_{v}-\theta_{u})} : \right.} \\
 \left.\theta_{v} \in \left[\frac{(m-1)\theta_{v}^{max}}{M}, \frac{m\theta_{v}^{max}}{M}\right], |\theta_{u}| \leq \frac{l\theta_{u}^{max}}{M}\right\}.
  \end{multline*}
  The weight on edge $(u_{l},v_{m})$ is equal to $\beta_{uv}(m,l) = \alpha(m,l)-\alpha(m,l+1)$ for $l=0,\ldots,(M-1)$ and $\beta_{uv}(m,M) = 0$.  By definition of $\alpha_{uv}(m,l)$, the weights $\beta_{uv}(m,l)$ are nonnegative.

We now define a class of subgraphs of $\tilde{G} = (\tilde{V},\tilde{E})$ that will be used to derive sufficient conditions for convergence.

\begin{definition}
\label{def:class_U_subgraph}
A subgraph $G^{\prime\prime} = (\tilde{V}, E^{\prime\prime})$ is a \emph{class-}$\mathcal{U}$ \emph{subgraph} if the neighbor set of each node $v_{m}$, denoted $N^{\prime\prime}(v_{m})$, satisfies
\begin{equation}
\label{eq:class_U_subgraph_def}
\sum_{u_{l} \in N^{\prime\prime}(v_{m})}{\beta_{uv}(m,l)} \geq \tau_{v}^{m},
\end{equation}
where $$\tau_{v}^{m} = \sum_{u \in N(v)}{\sum_{l=1}^{M}{\beta_{uv}(m,l)}} - \left(|\omega_{v}| - \sum_{u \in N(v)}{\alpha_{uv}(m,M)}\right).$$
\end{definition}

The following result gives a sufficient condition for convergence of $\boldsymbol{\theta}(t)$ to $\Lambda_{final}$ based on the connectivity of class-$\mathcal{U}$ subgraphs of $\tilde{G}$.

\begin{theorem}
\label{theorem:connectivity_convergence}
Suppose that a set $\Lambda_{PI} = \{\boldsymbol{\theta} : |\theta_{v}| \leq \theta_{v}^{PI}\}$ exists that is positive invariant and satisfies $\Lambda_{init} \subseteq \Lambda_{PI} \subseteq \Lambda_{bound}$.  Let $r(v) = \lceil \frac{M\gamma}{\theta_{v}^{max}}\rceil$ and $s(v) = \lceil \frac{M\theta_{v}^{PI}}{\theta_{v}^{max}}\rceil$. Define $V^{\prime} = \{v_{r(v)} : v \in V\}$ and $V^{\prime\prime} = \{v_{s(v)} : v \in V\}$.  If, for each node $v_{m} \in V^{\prime}$, there is a node $u_{l} \in V^{\prime\prime}$ and a directed path from $u_{l}$ to $v_{m}$, then for any $\boldsymbol{\theta}(0) \in \Lambda_{init}$, $\boldsymbol{\theta}(t) \in \Lambda_{final}$ for $t$ sufficiently large. %If there is a path from $V^{\prime\prime}$ to each node in $V^{\prime}$, then for any $\boldsymbol{\theta}_{0} \in \Lambda_{init}$, $\boldsymbol{\theta}(t) \in \Lambda_{final}$ for $t$ sufficiently large.
\end{theorem}

\begin{IEEEproof}
The approach of the proof will be to assume that $\boldsymbol{\theta}(t)$ does not converge to $\Lambda_{final}$, and then use Algorithm \ref{algo:sufficient_convergence} to construct a class-$\mathcal{U}$ subgraph in which $V^{\prime\prime}$ is not connected to $V^{\prime}$.  Define $\mathbf{X}^{\ast}$ to be the vector $\mathbf{X}[k]$ from Algorithm \ref{algo:sufficient_convergence} when the algorithm converges.

A subgraph $G^{\prime\prime}$ is constructed as follows.  For each node $v_{m}$ with $m < X_{v}^{\ast}$, define $N^{\prime\prime}(v_{m}) = \{u_{l} : u \in N(v), l=0,\ldots,X_{u}^{\ast}-1\}$.  For each node $v_{m}$ with $m \geq X_{v}^{\ast}$, select an arbitrary set $N^{\prime\prime}(v_{m})$ that satisfies (\ref{eq:class_U_subgraph_def}).

We now show that $G^{\prime\prime}$ is a class-$\mathcal{U}$ subgraph.  By construction, the sets $N^{\prime\prime}(v_{m})$ for $m \geq X_{v}^{\ast}$ satisfy (\ref{eq:class_U_subgraph_def}), so it remains to prove that Eq. (\ref{eq:class_U_subgraph_def}) holds for $\{v_{m} : m < X_{v}^{\ast}\}$.  For all such $v_{m}$, $$\sum_{u_{l} \in N^{\prime\prime}(v_{m})}{\beta_{uv}(m,l)} = \sum_{u \in N(v)}{\sum_{l=0}^{X_{u}^{\ast}-1}{\beta_{uv}(m,l)}}.$$ Now $N(v_{m}) = N^{\prime\prime}(v_{m}) \cup \{(u_{l},v_{m}) : l \geq X_{u}^{\ast}\}$ and this is a disjoint union.  Hence
\begin{IEEEeqnarray}{rCl}
\IEEEeqnarraymulticol{3}{l}{
\nonumber
\sum_{\stackrel{u_{l} \in}{N^{\prime\prime}(v_{m})}}{\beta_{uv}(m,l)}}\\
\nonumber
 &=& \sum_{u_{l} \in N(v_{m})}{\beta_{uv}(m,l)} - \sum_{u \in N(v)}{\sum_{l=X_{u}^{\ast}}^{M}{\beta_{uv}(m,l)}} \\
%\IEEEeqnarraymulticol{3}{l}{
\nonumber
&=& \sum_{u_{l} \in N(v_{m})}{\beta_{uv}(m,l)}  \\
\nonumber
&& - \sum_{u \in N(v)}{\sum_{l=X_{u}^{\ast}}^{M-1}{(\alpha_{uv}(m,l) - \alpha_{uv}(m,l+1))}} \\
\nonumber
&=& \sum_{u_{l} \in N(v_{m})}{\beta_{uv}(m,l)}\\
 \label{eq:submod_sufficient1}
&& - \sum_{u \in N(v)}{\left(\alpha_{uv}(m,X_{u}^{\ast}) - \alpha_{uv}(m,M)\right)}
\end{IEEEeqnarray}
Since Algorithm \ref{algo:sufficient_convergence} terminates with $X_{v}^{\ast} > m$, we must have $$\sum_{u \in N(v)}{\alpha_{uv}(m,X_{u}^{\ast})} < |\omega_{v}| + \epsilon.$$ Substituting this inequality into (\ref{eq:submod_sufficient1}) yields
\begin{multline*}
\sum_{u_{l} \in N^{\prime\prime}(v_{m})}{\beta_{uv}(m,l)} \\
> \sum_{u_{l} \in N(v_{m})}{\beta_{uv}(m,l)} - \left(|\omega_{v}| + \epsilon - \sum_{u \in N(v)}{\alpha_{uv}(m,M)}\right).
\end{multline*}
Hence $N^{\prime\prime}(v_{m})$ satisfies (\ref{eq:class_U_subgraph_def}), and $G^{\prime\prime}$ is a class-$\mathcal{U}$ subgraph.

By construction, there is no path from any node in $\{v_{m} : m \geq X_{v}^{\ast}\}$ to any node in $\{v_{m} : m < X_{v}^{\ast}\}$.  Since convergence to $\Lambda_{final}$ is not guaranteed, by Theorem \ref{theorem:algo_sufficient_convergence}, there exists $v$ with $r(v) < X_{v}^{\ast}$, and hence $V^{\prime} \cap \{v_{m} : m < X_{v}^{\ast}\} \neq \emptyset$.  At the same time, $V^{\prime\prime} \subseteq \{v_{m} : m \geq X_{v}^{\ast}\}$, since $X_{v}[0] = s(v)$ and $X_{v}$ is nonincreasing at each iteration of Algorithm \ref{algo:sufficient_convergence}.  Hence there exists at least one $v$ with $v_{r(v)}$ not path-connected to any node in $V^{\prime\prime}$, completing the proof.
\end{IEEEproof}

The following proposition established a converse result to Theorem \ref{theorem:connectivity_convergence}.  We show that, if the sufficient conditions defined for Algorithm \ref{algo:sufficient_convergence} in Theorem \ref{theorem:algo_sufficient_convergence} hold, then $V^{\prime\prime}$ is connected to $V^{\prime}$ in any class-$\mathcal{U}$ subgraph.
\begin{proposition}
\label{prop:aug_graph_algo}
If $X_{v}[k] = m$ in Algorithm \ref{algo:sufficient_convergence} at some iteration $k \geq 0$, then there exists a path from at least one node in $V^{\prime\prime}$ to $v_{m}$ in every class-$\mathcal{U}$ subgraph of $\tilde{G}$. %$v_{m}$ is connected to $V^{\prime\prime}$ in every class-$\mathcal{U}$ subgraph of $\tilde{G}$.
\end{proposition}

\begin{IEEEproof}
The proof is by induction on $k$.  By definition of the algorithm, $X_{v}[0] = s(v)$, and $v_{s(v)} \in V^{\prime\prime}$, implying that connectivity trivially holds when $k=0$.

Suppose that, at iteration $k$, node $v$ satisfies $X_{v}[k] = X_{v}[k-1]-1$.  Let $m=X_{v}[k]$   and define $G^{\prime\prime}$ to be a class-$\mathcal{U}$ subgraph, in which the neighbor set of $v_{m}$ is $N^{\prime\prime}(v_{m})$.  By definition of Algorithm \ref{algo:sufficient_convergence},
\begin{IEEEeqnarray*}{rCl}
|\omega_{v}| + \epsilon &<& \sum_{u \in N(v)}{\alpha_{uv}(m,X_{u}[k-1])} \\
\IEEEeqnarraymulticol{3}{l}{
= \sum_{u \in N(v)}{\left[\sum_{l=X_{u}[k-1]}^{M-1}{(\alpha_{uv}(m,l)-\alpha_{uv}(m,l+1))}\right.}} \\
&& \left. + \alpha_{uv}(m,M)\right]
\end{IEEEeqnarray*}
and rearranging terms yields $$\sum_{u \in N(v)}{\sum_{l=X_{u}[k-1]}^{M}{\beta_{uv}(m,l)}} > |\omega_{v}| + \epsilon - \sum_{u \in N(v)}{\alpha_{uv}(m,M)}.$$  We then have
\begin{IEEEeqnarray*}{rCl}
\IEEEeqnarraymulticol{3}{l}{
\sum_{u_{l} \in N^{\prime\prime}(v_{m})}{\beta_{uv}(m,l)} + \sum_{u \in N(v)}{\sum_{l=X_{u}[k-1]}^{M}{\beta_{uv}(m,l)}}} \\
&>& \sum_{u_{l} \in N(v_{m})}{\beta_{uv}(m,l)} - \left(|\omega_{v}| + \epsilon - \sum_{u \in N(v)}{\alpha_{uv}(m,M)}\right) \\
&& + |\omega_{v}| + \epsilon - \sum_{u \in N(v)}{\alpha_{uv}(m,M)} \\
&=& \sum_{u_{l} \in N(v_{m})}{\beta_{uv}(m,l)}.
\end{IEEEeqnarray*}
Since $$N^{\prime\prime}(v_{m}) \cup \{u_{l} : l \geq X_{u}[k-1]\} \subseteq N(v_{m})$$ and the weights $\beta_{uv}(m,l)$ are nonnegative, there exists $$u_{l} \in N^{\prime\prime}(v_{m}) \cap \{u_{l} : l \geq X_{u}[k-1]\}.$$  By induction, each node in $\{u_{l} : l \geq X_{u}[k-1]\}$ is connected to $V^{\prime\prime}$ in $G^{\prime\prime}$, and hence $v_{m}$ is connected to $V^{\prime\prime}$.
 %Then
%$$\sum_{u_{l} \in N^{\prime\prime}(v_{m})}{\beta_{uv}(m,l)} + \sum_{u \in N(v)}{\sum_{l=X_{u}[k-1]}^{M}{\beta_{uv}(m,l)}} >\sum_{u_{l} \in N(v_{m})}{\beta_{uv}(m,l)},$$ implying that $$N^{\prime\prime}(v_{m}) \cap \{u_{l} : l \geq X_{u}[k-1]\} \neq \emptyset$$ since the weights $\beta_{uv}(m,l)$ are nonnegative.  Each node $u_{l}$ satisfying $l \geq X_{u}[k-1]$ is connected to $V^{\prime\prime}$ by inductive hypothesis, and hence $v_{m}$ is connected to $V^{\prime\prime}$ as well.
\end{IEEEproof}

Proposition \ref{prop:aug_graph_algo} leads to the following bound on the function $g(A)$ introduced in Section \ref{subsec:sync}.

\begin{corollary}
\label{corollary:g_convergence}
The function $g(A)$ satisfies
\begin{multline}
\label{eq:g_convergence}
g(A) \leq \max_{v}{\left\{\min_{m}{\left\{\frac{m\theta_{v}^{max}}{M} : v_{m} \mbox{ connected to } V^{\prime\prime}\right.}\right.} \\
 \left. \left.\mbox{ in all class-}\mathcal{U} \mbox{ subgraphs}\right\}\right\}.
\end{multline}
\end{corollary}

\begin{IEEEproof}
By Proposition \ref{prop:aug_graph_algo}, if $v_{m}$ is connected to $V^{\prime\prime}$ in all class-$\mathcal{U}$ subgraphs, then $|\theta_{v}^{\ast}| \leq \frac{m\theta_{v}^{max}}{M}$ for any equilibrium point $\boldsymbol{\theta}^{\ast}$.  Eq. (\ref{eq:g_convergence}) follows from the fact that $g(A)$ is defined as the largest value of $\gamma$ such that $|\theta_{v}^{\ast}| = \gamma$ for some equilibrium $\boldsymbol{\theta}^{\ast}$ and node $v$.
\end{IEEEproof}

We now show that convergence to $\Lambda_{final}$ can be mapped to a submodular constraint on the set of input nodes $A$.  Let $\pi$ be a probability distribution on the set of class-$\mathcal{U}$ subgraphs with $\pi(G^{\prime}) > 0$ for each class-$\mathcal{U}$ subgraph $G^{\prime}$.  Define the function $h_{2}(A)$ as $$h_{2}(A) \triangleq \sum_{G^{\prime} \in \mathcal{U}}{\pi(G^{\prime})f(A|G^{\prime})},$$ where $f(A|G^{\prime})$ is defined by
\begin{multline*}
f(A|G^{\prime}) = |\left\{v_{m} \in V^{\prime} : \mbox{there exists } u_{l} \in V^{\prime\prime}\right.\\
 \left.\mbox{ and a path from $u_{l}$ to $v_{m}$}\right\}|.
 \end{multline*}
 The function $h_{2}(A)$ can be approximated by randomly sampling a collection of class-$\mathcal{U}$ subgraphs and computing $f(A|G^{\prime})$ for each subgraph.  The following corollary describes convergence to $\Lambda_{final}$ based on $h_{2}(A)$.%is the number of nodes in $V^{\prime}$ that are connected to $V^{\prime\prime}$ in $G^{\prime}$.

\begin{corollary}
\label{corollary:convergence}
If $h_{2}(A) = n$, then for any $\boldsymbol{\theta}(0) \in \Lambda_{init}$, $\boldsymbol{\theta}(t) \in \Lambda_{final}$ for $t$ sufficiently large.
\end{corollary}

\begin{IEEEproof}
We have that $f(A|G^{\prime}) \leq n$ and the function $h_{2}(A)$ is a convex combination of the values of $f(A|G^{\prime})$.  Hence $h_{2}(A) \leq n$ with equality if and only if $f(A|G^{\prime}) = n$ for all class-$\mathcal{U}$ subgraphs $G^{\prime}$.
By definition of $f(A|G^{\prime})$, for every node $v_{m} \in V^{\prime}$, there exists a path from $V^{\prime\prime}$ to $v_{m}$ in each class-$\mathcal{U}$ subgraph.  Convergence to $\Lambda_{final}$ is then implied by Theorem \ref{theorem:connectivity_convergence}.
\end{IEEEproof}

Finally, to show that $h_{2}(A) = n$ is a submodular constraint in $A$, we have the following result establishing submodularity of $h_{2}(A)$.
\begin{proposition}
\label{prop:convergence_submodular}
The function $h_{2}(A)$ is submodular as a function of $A$.
\end{proposition}

\begin{IEEEproof}
We first show that $f(A|G^{\prime})$ is submodular as a function of $A$.  The impact of adding a node $\{v\}$ to $A$ is to increase the set $V^{\prime\prime}$ to $V^{\prime\prime} \cup \{v_{m} : m \geq 0\}$.  Hence, the increment $f(A \cup \{v\}) - f(A)$ is equal to the number of nodes in $V^{\prime}$ that are connected to $\{v_{m} : m \geq 0\}$ but are not connected to $V^{\prime\prime}$.  If $A \subseteq B$, then any node that is not connected to $V^{\prime\prime}$ with input set $B$ is automatically not connected to $V^{\prime\prime}$ with input set $A$.  Hence $$f(A \cup \{v\} | G^{\prime}) - f(A|G^{\prime}) \geq f(B \cup \{v\}|G^{\prime}) - f(B|G^{\prime}),$$ establishing submodularity of $f(A|G^{\prime})$.  The function $h_{2}(A)$ is equal to a nonnegative weighted sum of submodular functions, and hence is submodular.  %For any sets $A \subseteq B$ and any $v \notin B$, we have that the increment $f(A \cup \{v\}) - f(A)$ is equal to the set of nodes that are connected to $v_{0}$ in $G^{\prime}$ but are not connected to $V^{\prime}$ when $v \notin A$.     Hence the increment property $$f(A \cup \{v\}|G^{\prime}) - f(A|G^{\prime}) \geq f(B \cup \{v\}|G^{\prime}) - f(B|G^{\prime})$$ follows.  The function $h_{2}(A)$ is therefore a nonnegative weighted sum of submodular functions, and hence is submodular.
\end{IEEEproof}

Proposition \ref{prop:convergence_submodular} implies that convergence to $\Lambda_{final}$ can be formulated as a submodular constraint $h_{2}(A) = n$, which ensures that all nodes in $V^{\prime}$ are connected to $V^{\prime\prime}$.  In the following section, we demonstrate how to combine the submodular constraints of Sections \ref{subsec:submod_PI} and \ref{subsec:submod_convergence} to formulate selection of input nodes for practical synchronization as a submodular optimization problem.

 \subsection{Submodular Optimization Approach to Selecting Inputs}
 \label{subsec:input_selection}
 The conditions derived in Sections \ref{subsec:submod_PI} and \ref{subsec:submod_convergence} imply that the nodes achieve practical synchronization if $h_{1}(A|G) = 0$ for some class-$\mathcal{T}$ subgraph $G$ and $h_{2}(A) = n$.   %we have that the nodes achieve practical node synchronization if $r_{1}(A) + r_{2}(A) = 2n$ and practical edge synchronization if $r_{1}^{\prime}(A) + r_{2}^{\prime}(A) = 2p$.
 We consider two input selection problems.   In the first problem, we select the minimum-size set of input nodes in order to guarantee $\gamma$-practical synchronization.  In the second problem, we select a set of up to $k$ input nodes in order to maximize the input cohesiveness $g(A)$.

 The problem of selecting the minimum-size input nodes in order to guarantee $\gamma$-practical synchronization can be formulated as
 \begin{equation}
 \label{eq:min_input_sync}
 \begin{array}{ll}
 \mbox{minimize} & |S| \\
 \mbox{s.t.} & h_{1}(A|G^{\prime}) = 0 \ \mbox{for some class-$\mathcal{T}$ subgraph $G^{\prime}$} \\
  & h_{2}(A) = n
  \end{array}
  \end{equation}

 In solving (\ref{eq:min_input_sync}), we first observe that the function $h_{2}(A)$ depends on the positive invariant set $\Lambda_{PI}$.  Hence, we propose a sequential approach, in which a set of positive invariant sets, and the minimum-size set of input nodes required to ensure positive invariance, are computed.  A minimum-size set of additional input nodes required to ensure convergence to $\Lambda_{final}$ is then computed.

 The first step in the procedure is to generate a random collection of class-$\mathcal{T}$ subgraphs $\{G_{1}^{\prime},\ldots,G_{N}^{\prime}\}$.  For each subgraph, the optimization problem
 \begin{equation}
 \label{eq:first_stage_opt}
 \mbox{minimize}\{|A| : h_{1}(A|G_{i}) = 0\}
 \end{equation}
 is approximated.  One for solving (\ref{eq:first_stage_opt}) is as follows.  Initialize a set $A_{i} = \emptyset$.  At each iteration, select a node $\{v\}$ such that $h(A_{i}) - h(A_{i} \cup \{v\})$ is maximized, terminating when $h(A_{i}) = 0$.

 After obtaining sets $A_{1},\ldots,A_{N}$, the next step in the procedure is to ensure that the condition $h_{2}(A) = n$ is satisfied.  This condition can be ensured by solving
 \begin{equation}
 \label{eq:second_stage_opt}
 \mbox{minimize}\{|A| : h_{2}(A \cup A_{i}) = n\}
 \end{equation}
 for $i=1,\ldots,N$, resulting in a collection $A_{1}^{\prime},\ldots,A_{N}^{\prime}$.  The set $A_{j}^{\prime}$ with minimum cardinality is then chosen as the minimum-size set for guaranteeing practical synchronization.  This procedure is shown as Algorithm \ref{algo:minimum_input_PS}.

   \begin{center}
\begin{algorithm}[!htp]
	\caption{Algorithm for selecting the minimum-size input set $A$ to guarantee practical synchronization.}
	\label{algo:minimum_input_PS}
	\begin{algorithmic}[1]
		\Procedure{Min\_Input\_PS}{$G=(V,E)$, $\omega_{1},\ldots,\omega_{n}$, $\gamma$, $\{\theta_{1}^{max}, \ldots, \theta_{n}^{max}\}$, $\{\theta_{1}^{0},\ldots,\theta_{n}^{0}\}$}
            \State \textbf{Input}: Graph $G=(V,E)$, intrinsic frequencies $\omega_{1},\ldots,\omega_{n}$, bound $\gamma$, boundary values $\{\theta_{i}^{max} : i \in V\}$, initial states $\{\theta_{i}^{0} : i \in V\}$.
            \State \textbf{Output}: Set of input nodes $A$
            \State Generate a collection of class-$\mathcal{T}$ subgraphs $G_{1},\ldots,G_{N}$
            \For{$i=1,\ldots,N$}
            \State $A_{i} \leftarrow \emptyset$
            \While{$h_{1}(A_{i}|G_{i}^{\prime}) > 0$}
            \State $v \leftarrow \arg\min{\left\{h_{1}(A_{i} \cup \{v\} | G_{i}^{\prime}) : v \in V\right\}}$
            \State $A_{i}^{\prime} \leftarrow \emptyset$
            \EndWhile
            \State $(\overline{\theta}_{1}^{(i)},\ldots,\overline{\theta}_{n}^{(i)}) \leftarrow$ $\mbox{Identify\_PI\_Set}(G,$ $(\omega_{1},\ldots,\omega_{n}),$  $A_{i}^{\prime}, \{\theta_{v}^{0} : v \in V\}, \{\theta_{v}^{max} : v\in V\})$
            \State $A_{i}^{\prime} \leftarrow \emptyset$
            \While{$h_{2}(A_{i} \cup A_{i}^{\prime}) < n$}
            %\For{$v \in V \setminus A$}
            \State $v \leftarrow \arg\max{\left\{h_{2}(A_{i} \cup A_{i}^{\prime} \cup \{v\}): v \in V\right\}}$
            \State $A_{i}^{\prime} \leftarrow A_{i}^{\prime} \cup \{v\}$
            \EndWhile
            \EndFor
            \State $j \leftarrow \arg\min{\{|A_{i} \cup A_{i}^{\prime}| : i=1,\ldots,N\}}$
            \State $A \leftarrow A_{j} \cup A_{j}^{\prime}$; \Return{$A$}
		\EndProcedure
	 \end{algorithmic}
\end{algorithm}
\end{center}

\begin{lemma}
\label{lemma:dual_correct}
Algorithm \ref{algo:minimum_input_PS} returns a set $A$ that guarantees practical synchronization.
\end{lemma}

\begin{IEEEproof}
The set $A$ returned by Algorithm \ref{algo:minimum_input_PS} can be decomposed as $A = A_{i} \cup A_{i}^{\prime}$ for some index $i$.  The set $A_{i}$ satisfies $h_{1}(A_{i} |G_{i}) = 0$ for the class-$\mathcal{T}$ subgraph $G_{i}$, and hence there exists a positive invariant subset $\Lambda_{PI}$ satisfying $\Lambda_{init} \subseteq \Lambda_{PI} \subseteq \Lambda_{bound}$ when the input set is $A$.  Furthermore, $h_{2}(A_{i} \cup A_{i}^{\prime}) = n$, and hence convergence to $\Lambda_{final}$ is guaranteed by Theorem \ref{theorem:connectivity_convergence}.  Combining these two conditions leads to guaranteed practical synchronization by Theorem \ref{theorem:sufficient_practical_sync}.
\end{IEEEproof}

We next analyze the optimality guarantees provided by this approach.

\begin{lemma}
\label{lemma:dual_optimality_guarantees}
For each $i=1,\ldots,N$, the sets $A_{i}$ and $A_{i}^{\prime}$ selected by Algorithm \ref{algo:minimum_input_PS} satisfy
\begin{IEEEeqnarray}{rCl}
\label{eq:first_dual_bound}
\frac{|A_{i}|}{|A_{i}^{\ast}|} &\leq& 1 + \ln{n} \\
\label{eq:second_dual_bound}
\frac{|A_{i}^{\prime}|}{A_{i}^{\prime \ast}} &\leq& 1 + \ln{\left\{\frac{n}{h_{2}(A_{i} \cup A_{i}^{\prime}) - h_{2}(A_{i} \cup \hat{A}_{i}^{\prime})}\right\}}
\end{IEEEeqnarray}
where $A_{i}^{\ast}$ and $A_{i}^{\prime \ast}$ denote the optimal solutions to (\ref{eq:first_stage_opt}) and (\ref{eq:second_stage_opt}), respectively and $\hat{A}_{i}^{\prime}$ is the value of $A_{i}^{\prime}$ at the second-to-last iteration of the second inner loop in Algorithm \ref{algo:minimum_input_PS}.
\end{lemma}

\begin{IEEEproof}
Problems (\ref{eq:first_stage_opt}) and (\ref{eq:second_stage_opt}) involve minimizing the cardinality of a set subject to a submodular constraint.  For the problem of minimizing $|A|$ subject to a constraint that $\{f(A) \leq \alpha\}$ for some $\alpha$, the greedy algorithm is known to return a set $A$ satisfying $$\frac{|A|}{|A^{\ast}|} \leq 1 + \ln{\left\{\frac{\alpha}{f(A) - f(\hat{A})}\right\}},$$ where $\hat{A}$ is the set computed by the greedy algorithm at the second-to-last iteration.  This bound reduces to (\ref{eq:first_dual_bound}) and (\ref{eq:second_dual_bound}) for the problems (\ref{eq:first_stage_opt}) and (\ref{eq:second_stage_opt}), respectively.
\end{IEEEproof}

The runtime of Algorithm \ref{algo:minimum_input_PS} is determined by the network size $n$ and the number of class-$\mathcal{T}$ subgraphs generated, and is bounded by $O(n^{2}N)$ evaluations of $h_{1}(A|G^{\prime})$ and $h_{2}(A)$.

 \begin{figure*}
\centering
$\begin{array}{ccc}
\includegraphics[width=2.3in]{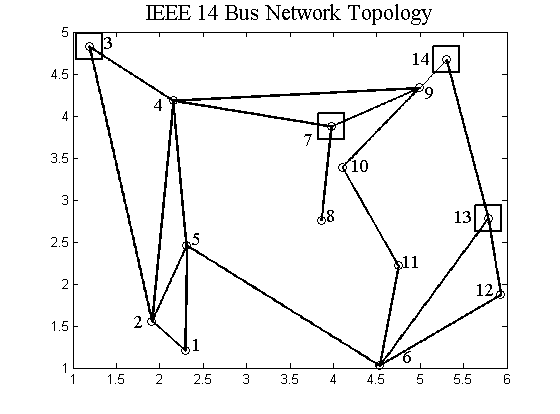} &
\includegraphics[width=2.3in]{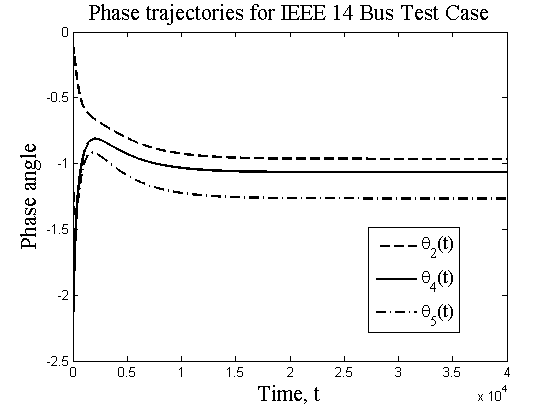} &
\includegraphics[width=2.3in]{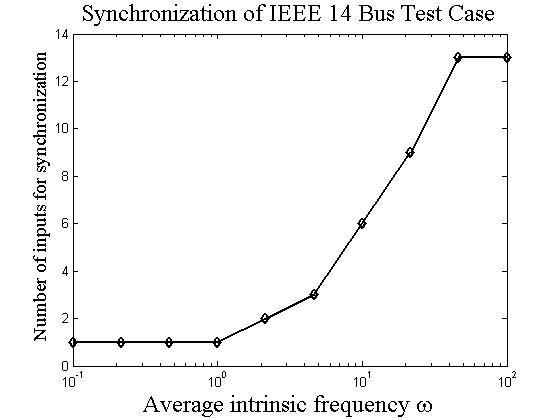} \\
\mbox{(a)} & \mbox{(b)} & \mbox{(c)}
\end{array}$
\caption{Synchronization case study using the IEEE 14 Bus case study \cite{14bus}.  Coupling coefficients were chosen to be equal to the inverse of the magnitudes of the line impedances, while the intrinsic frequencies were chosen from a Gaussian distribution.  (a) Illustration of 14-bus network and inputs needed for one trial when the variance of the intrinsic frequencies was equal to $8$.  The black squares indicate input nodes.  (b) Sample trajectories for nodes $2$, $4$, and $5$ in the 14-bus network.  Node trajectories converge to within the desired phase difference ($\gamma = \pi/3$) of each other. (c) Number of inputs required for practical edge synchronization as a function of the variance of the intrinsic frequencies.  As the variance increases, the frequencies of neighboring nodes diverge and hence more inputs are needed to ensure synchronization.}
\label{fig:power_simulation}
\end{figure*}

We now consider the problem of selecting up to $k$ input nodes in order to maximize the input cohesiveness $g(A)$, formulated as
\begin{equation}
\label{eq:primal}
\begin{array}{ll}
\mbox{maximize} & g(A) \\
\mbox{s.t.} & |A| \leq k \\
 & h_{1}(A|G^{\prime}) = 0 \ \mbox{for some class-$\mathcal{T}$ subgraph $G^{\prime}$}
 \end{array}
 \end{equation}
In order to approximate the solution to (\ref{eq:primal}),  we define the function $h_{3}(A)$ to be the expected number of nodes connected to $V^{\prime\prime}$ in a random class-$\mathcal{U}$ subgraph, i.e.,
\begin{multline*}
h_{3}(A) = \mathbf{E}\left(|\left\{v_{m} : \mbox{ there exists a node } u_{l} \in V^{\prime\prime}\right. \right.\\
 \left. \left.\mbox{ connected to } v_{m}\right\}|\right).
\end{multline*}
The definition of $h_{3}(A)$ is motivated by Corollary \ref{corollary:g_convergence}, which establishes an upper bound on $g(A)$ based on the number of nodes that are connected to $V^{\prime\prime}$.

By the same argument as Proposition \ref{prop:convergence_submodular}, $h_{3}(A)$ is submodular as a function of $A$.  We consider the problem of maximizing $h_{3}(A)$, subject to the constraint that $h_{2}(A|G^{\prime}) = 0$ for some class-$\mathcal{T}$ subgraph $G^{\prime}$ and $|A| \leq k$.  Our approach is a sequential algorithm in which a collection of subsets $A_{1},\ldots,A_{N}$ is found that guarantee the existence of a positive invariant set in a collection of class-$\mathcal{T}$ subgraphs $G_{1}^{\prime},\ldots,G_{N}^{\prime}$.  For each subset of $A_{i}$, we then select a subset $A_{i}^{\prime}$ such that $|A_{i} \cup A_{i}^{\prime}| \leq k$ and $h_{3}(A_{i} \cup A_{i}^{\prime})$ is maximized.  A formal description is given as Algorithm \ref{algo:maximize_PS}.

   \begin{center}
\begin{algorithm}[!htp]
	\caption{Algorithm for selecting the minimum-size input set $A$ to guarantee practical synchronization.}
	\label{algo:maximize_PS}
	\begin{algorithmic}[1]
		\Procedure{Maximize\_Sync}{$G=(V,E)$, $\omega_{1},\ldots,\omega_{n}$, $k$, $\{\theta_{1}^{max}, \ldots, \theta_{n}^{max}\}$, $\{\theta_{1}^{0},\ldots,\theta_{n}^{0}\}$}
            \State \textbf{Input}: Graph $G=(V,E)$, intrinsic frequencies $\omega_{1},\ldots,\omega_{n}$, number of input nodes $k$, boundary values $\{\theta_{i}^{max} : i \in V\}$, initial states $\{\theta_{i}^{0} : i \in V\}$.
            \State \textbf{Output}: Set of input nodes $A$
            \State Generate a collection of class-$\mathcal{T}$ subgraphs $G_{1},\ldots,G_{N}$
            \For{$i=1,\ldots,N$}
            \State $A_{i} \leftarrow \emptyset$
            \While{$h_{1}(A_{i}|G_{i}^{\prime}) > 0$}
            \State $v \leftarrow \arg\min{\left\{ h_{1}(A_{i} \cup \{v\} | G_{i}^{\prime}) : v \in V\right\}}$
            \State $A_{i}^{\prime} \leftarrow \emptyset$
            \EndWhile
            \State $(\overline{\theta}_{1}^{(i)},\ldots,\overline{\theta}_{n}^{(i)}) \leftarrow$ $\mbox{Identify\_PI\_Set}(G,$ $(\omega_{1},\ldots,\omega_{n}),$  $A_{i}^{\prime}, \{\theta_{v}^{0} : v \in V\}, \{\theta_{v}^{max} : v\in V\})$
            \State $A_{i}^{\prime} \leftarrow \emptyset$
            \If{$|A_{i}| < k$}
            \While{$|A_{i} \cup A_{i}^{\prime}| \leq k$}
            \State $v \leftarrow \arg\max{\left\{h_{3}(A_{i} \cup A_{i}^{\prime} \cup \{v\}): v \in V\right\}}$
            \State $A_{i}^{\prime} \leftarrow A_{i}^{\prime} \cup \{v\}$
            \EndWhile
            \EndIf
            \EndFor
            \State $j \leftarrow \arg\max{\{h_{3}(A_{i} \cup A_{i}^{\prime}) : |A_{i}| \leq k\}}$
            \State $A \leftarrow A_{j} \cup A_{j}^{\prime}$; \Return{$A$}
		\EndProcedure
	 \end{algorithmic}
\end{algorithm}
\end{center}

The following lemma describes the optimality guarantees of this approach.

\begin{lemma}
\label{lemma:primal_optimality}
Define $$A_{i}^{\prime\ast} = \arg\max_{A_{i}^{\prime}}{\{h_{3}(A_{i} \cup A_{i}^{\prime}) : |A_{i} \cup A_{i}^{\prime}| \leq k\}},$$ and let $A$ denote the set returned by Algorithm \ref{algo:maximize_PS}.  Then
\begin{equation}
\label{eq:primal_bound}
h_{3}(A) \geq \left(1-1/e\right)h_{3}(A_{i} \cup A_{i}^{\prime\ast}).
\end{equation}
\end{lemma}

\begin{IEEEproof}
For the problem of selecting a set $A$ of cardinality $k$ to maximize a monotone submodular objective function, the greedy algorithm is known to provide an optimality bound of $(1-1/e)$.  The problem of maximizing $h_{3}(A)$ subject to a constraint on $k$ has this structure, leading to the optimality bound (\ref{eq:primal_bound}).
\end{IEEEproof}

In this case, the runtime is $O(knN)$ evaluations of $h_{1}(A|G^{\prime})$ and $h_{3}(A)$.  Hence the problem of selecting a minimum-size set of input nodes to guarantee practical synchronization (Eq. (\ref{eq:min_input_sync})), as well as the problem of selecting a fixed-size set of input nodes to maximize input cohesiveness (Eq. (\ref{eq:primal})), can be approximated within the submodular optimization framework.

\section{Numerical Results}
\label{sec:simulation}
We investigated our approach through a numerical study.  The goal of our study was to analyze the impact of the external inputs on the oscillator dynamics, as well as to investigate the properties of the inputs chosen by our algorithm. We  considered synchronization in three application domains: (i) synchronization of phase angles in the IEEE 14 Bus power system test case \cite{14bus}, (ii) time synchronization in wireless networks~\cite{baldoni2010coupling}, modeled as a geometric random graph,  and (iii) synchronization of firing rates in the \emph{C. Elegans} neuronal network \cite{varshney2011structural}.  The results of each study are summarized as follows.  In all cases, the parameters $M=N=20$.

\begin{figure*}
\centering
$\begin{array}{ccc}
\includegraphics[width=2.3in]{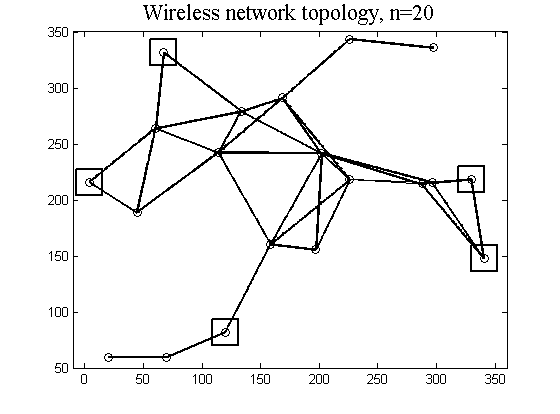} &
\includegraphics[width=2.3in]{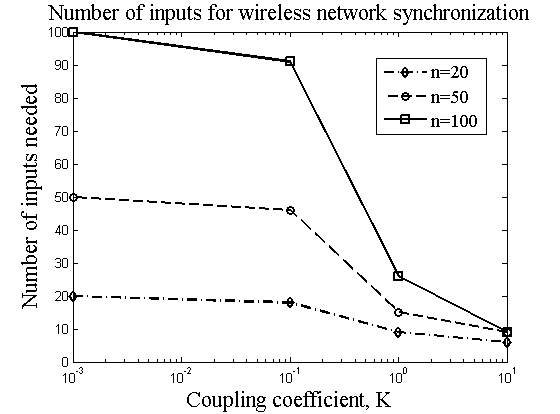} &
\includegraphics[width=2.3in]{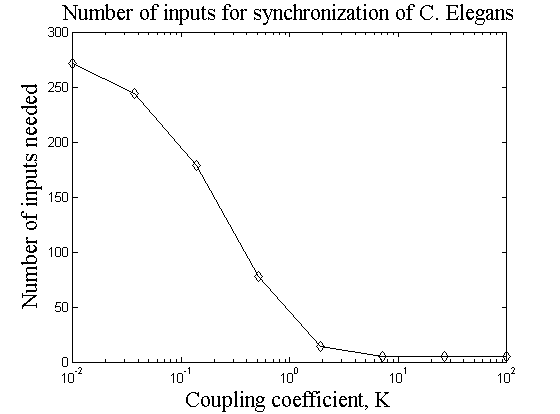} \\
\mbox{(a)} & \mbox{(b)} & \mbox{(c)}
\end{array}$
\caption{Numerical results for time synchronization and neuronal synchronization.  (a) Illustration of inputs chosen for a wireless network, modeled as a geometric random graph, when $K=0.7197$, $n=20$, and the input nodes are indicated by squares.  The set of input nodes is based on the network topology and intrinsic frequencies of the nodes. (b) Number of input nodes required for time synchronization.  The number of nodes is a decreasing function of the coupling coefficient.  The reduction in the fraction of input nodes required for synchronization is most significant for larger network size. (c) Number of input nodes required for synchronization of the \emph{C. Elegans} network ($n=279$) when the intrinsic frequencies have Gaussian distribution with unit variance.  The number of inputs is a decreasing function of the coupling coefficient.}
\label{fig:other_sim}
\end{figure*}

\subsection{Phase Synchronization in Power System}
\label{subsec:power_system_sim}
 Synchronization plays a vital role in the power system, in which stable operation requires buses and generators to maintain the same frequency and a relative phase difference of no more than $\pi/2$ on each transmission line (edge) \cite{bergen2009power}.  In order to evaluate our approach to synchronization of power systems, we consider the first-order model (\ref{eq:KM}) with negligible resistance on the transmission lines.   The input nodes represent generators that are fixed to a common reference phase and frequency in order to restore stability to the grid.
We studied phase synchronization in the power system using the IEEE 14 Bus case study \cite{14bus}, which provides a network topology with $n=14$ nodes along with the line impedances of the edges.  We defined the coupling coefficient $K_{uv} = 1/\eta_{uv}$, where $\eta_{uv}$ is the magnitude of the impedance of edge $(u,v)$.  The intrinsic frequencies were chosen according to a zero-mean Gaussian distribution with variance ranging from $0.01$ to $100$ in different trials.  The goal was to ensure that the phase angles satisfied $|\theta_{v} - \theta_{u}| < \frac{\pi}{3}$ (i.e., $\gamma$-practical edge synchronization with $\gamma = \pi/3$), while ensuring that $|\theta_{v}(t)| \leq \pi/4$ for all $t$, consistent with the goal of guaranteeing power system stability \cite{bergen2009power}.

Figure \ref{fig:power_simulation}(a) illustrates input selection for a given set of intrinsic frequencies.  The variance of the intrinsic frequencies was equal to $8$.  The dark squares indicate input nodes selected by our algorithm ($|A| = 4$).  We observe that each non-input node is at most two hops away from an input node, which suggests that centrally located nodes are more likely to be candidates for inputs.  On the other hand, nodes with high degree were not necessarily chosen as inputs.  Additional information beyond the network topology, including the intrinsic frequencies and the coupling coefficients between nodes, is incorporated into the input selection process.

In Figure \ref{fig:power_simulation}(b), trajectories of non-input nodes $2$, $4$, and $5$ from Figure \ref{fig:power_simulation}(a) are shown.  The trajectories of these neighboring nodes converge to within the desired bound of $\frac{\pi}{3}$ from each other, and synchronize to the frequency of the input nodes ($\omega_{0} = 0$; by Lemma \ref{lemma:frame_of_reference}, an arbitrary initial frequency could be chosen, but we selected $\omega_{0} = 0$ for clarity of the figure).  %Practical edge synchronization occurs in spite of the fact that the initial phases of neighboring nodes $2$ and $5$ differed by more than $\pi/2$.

Figure \ref{fig:power_simulation}(c) shows the number of input nodes required to achieve practical synchronization for different intrinsic frequencies.  When the variance of $\{\omega_{v} : v \in V\}$ is low, only one input node is required to achieve synchronization.  For variance near 10, an intermediate number of inputs is required for synchronization, while for large variance in the intrinsic frequencies all nodes must be selected as input nodes to guarantee synchronization.

\subsection{Time Synchronization in Wireless Network}
\label{subsec:time_sync_sim}
We analyzed time synchronization in wireless networks, in which the input nodes represent anchor nodes with clocks that are set to the reference time. We modeled the wireless network as a geometric random graph.  Each node was assumed to share an edge with other nodes within a range of $100$m.  The simulation was run with the number of nodes $n = \{20,50,100\}$.  The nodes were assumed to be deployed uniformly at random over a square region, with area chosen so that each node had three neighbors on average.  Nodes were assumed to have i.i.d. Gaussian intrinsic frequency with unit variance and zero mean.  We computed the number of input nodes required to ensure $\gamma$-practical node synchronization with $\gamma = \pi/5$ of each other.  The edges had identical coupling with $K_{uv} = K$ for all $(u,v) \in E$, with $K$ in the set $\{0.01, 0.01, 0.1, 1, 10\}$.

Figure \ref{fig:other_sim}(a) shows one example of a geometric random graph with $n=20$ and $K=0.7197$ for all links.  The input nodes are indicated by black squares.  As in the power system case study, most non-input nodes are within two hops of an input node, and high-degree nodes are not necessarily chosen as inputs.  Note that two adjacent nodes are both chosen as inputs, due to the high intrinsic frequencies of those nodes.

The number of nodes required for synchronization, shown in Figure \ref{fig:other_sim}(b), is a decreasing function of the coupling coefficient, since a stronger coupling implies that input nodes are able to steer their neighbors towards synchronization.  We observe that the fraction of input nodes required for synchronization is smaller for larger networks.

\subsection{Firing Rate Synchronization in Neuronal Network}
\label{subsec:neuronal_sim}
Synchronization of groups of neuronal cells to a common frequency plays a role in information storage and processing for motor control \cite{cassidy2002movement} and memory \cite{klimesch1996memory}.  We analyzed the impact of input nodes on synchronization using the \emph{Caenorhabditis Elegans} (roundworm) neuronal network \cite{varshney2011structural}.  The number of nodes is $n=279$.  The intrinsic frequencies were independently from a zero-mean, unit-variance Gaussian distribution.  All links had the same coupling coefficient $K$, which varied from $K=0.001$ to $K=10$.  The goal was to achieve practical node synchronization with  $\gamma = \pi/4$.

We observed that the number of input nodes required decreases as the coupling coefficient increases.  The reduction in the number of nodes required was larger than in the random geometric graph (Section \ref{subsec:time_sync_sim}), even though the neuronal network contained more nodes.  An explanation for this phenomenon is that the neuronal network has higher degree, as well as additional links between different regions of the network.  This more connected network topology enables synchronization of more nodes and at a lower coupling coefficient. 
\section{Conclusions and Future Work}
\label{sec:conclusion}
We considered the problem of ensuring practical synchronization of phase-coupled oscillators from any initial state by introducing external input nodes.  We studied two synchronization problems.  In the first problem, the goal is for all node phases to converge to within a bound of a given reference phase, while also converging to the same frequency.  In the second problem, the relative phase differences between neighboring nodes must converge to a given bound, while all nodes converge to the same frequency.

We derived sufficient conditions for achieving both synchronization goals.  We proved that practical synchronization is satisfied if two conditions hold.  First, if there exists a positive invariant set $\Lambda_{PI}$ that contains the set of initial node states and satisfies $\Lambda_{PI} \subseteq \Lambda_{bound}$, where $\Lambda_{bound}$ defines the set of upper bounds on the node phases.  Second, for any initial state inside $\Lambda_{PI}$, each node's phase is guaranteed to converge to within the desired level of synchronization.  We developed efficient algorithms for verifying both conditions. %We proved that each synchronization criterion is achieved if there exists at least one equilibrium that lies within the desired region (either all nodes within a given bound of the reference phase, or all relative differences within a given bound), and no stable equilibria outside the desired region.  We then presented a threshold-based condition for existence of an equilibrium within the desired region, as well as an efficient algorithm for verifying existence of such an equilibrium.  Finally, we derived a class of conditions for non-existence of stable equilibria outside the desired region, and proposed an algorithm for verifying non-existence of undesired equilibria for a given set of input nodes.

We formulated a submodular optimization approach for selecting a set of input nodes to guarantee synchronization.  To achieve a submodular formulation, we first constructed an augmented network graph.  We identified two classes of subgraphs (denoted class-$\mathcal{T}$ and class-$\mathcal{U}$) of the augmented graph.  We proved that the sufficient condition for existence of a positive invariant set is equivalent to the existence of a cut in at least one class-$\mathcal{T}$ subgraph.  We then showed that the sufficient condition for convergence to practical synchronization is equivalent to a connectivity constraint for over all class-$\mathcal{U}$ subgraphs.  %existence of a desired equilibrium is equivalent to connectivity between each non-input node and the input set in all class-$\mathcal{T}$ subgraphs.  We then showed that the sufficient conditions for non-existence of undesired equilibria are equivalent to connectivity between each non-input node and the input set in all class-$\mathcal{U}$ subgraphs.

 We  mapped this condition to submodular constraints on the set of input nodes.  Based on the submodular formulation, we proposed efficient algorithms with provable optimality bounds for selecting a set of up to $k$ nodes  in order to maximize the level of synchronization in the network, as well as selecting the minimum-size input set to guarantee a desired level of synchronization. %We demonstrated that the sufficient conditions for existence of a desired equilibrium and non-existence of undesired equilibria can be equivalently formulated as submodular constraints on the set of input nodes.  Our approach was to derive equivalent conditions based on the connectivity of an augmented network graph.

Our approach was validated through a numerical study of synchronization in power system, wireless, and neuronal networks.  Our numerical study supports the intuition that the number of input nodes required decreases as the coupling between neighboring nodes grows stronger.  In addition, we found that centrally located nodes are often chosen as inputs, so that the maximum distance between a node and the input set is small.

The threshold-based conditions derived in this work are sufficient, but not necessary.  Characterizing the space of networks where these conditions are also necessary, as well as developing tighter sufficient conditions, remains an open problem.  Furthermore, while the model we studied considers the first-order dynamics of the nodes, second-order dynamics are often used to model systems such as the power grid \cite{dorfler2013synchronization}.  Generalizing our approach to these second-order systems is a direction of future research. 

\bibliographystyle{IEEEtran}
\bibliography{TAC14}

\end{document}